\documentclass{aa}

\usepackage[english]{babel}
\usepackage{graphicx}
\usepackage{newlfont}
\usepackage{amssymb}

\usepackage{amsmath}
\usepackage{latexsym}
\usepackage{amsthm}
\usepackage{natbib}
\bibpunct{(}{)}{;}{a}{}{,} 

\usepackage{siunitx}
\DeclareSIUnit\parsec{pc}
\DeclareSIUnit\erg{erg}
\DeclareSIUnit\deg{deg^{-2}}
\DeclareSIUnit\count{cnt}

\begin{document}
\title{Chandra ACIS-I particle background: an analytical model}
\author{I. Bartalucci\inst{\ref{inst1}}
\and P. Mazzotta\inst{\ref{inst1},\ref{inst2}} 
\and H. Bourdin\inst{\ref{inst1}} 
\and A. Vikhlinin\inst{\ref{inst2}}
\institute{Department of Physics, Universit\`a di Roma “Tor Vergata”, via della Ricerca Scientifica 1, 00133 Rome, Italy\label{inst1} 
\and Harvard-Smithsonian Center for Astrophysics, 60 Garden St, Cambridge, MA 02138, USA\label{inst2}}}

\abstract{}{Imaging and spectroscopy of X-ray extended sources require a proper characterisation of a spatially unresolved background signal.
This background includes sky and instrumental components, each of which are characterised by its proper spatial and spectral 
behaviour. While the X-ray sky background has been extensively studied in previous work, here we analyse and model the instrumental 
background of the ACIS-I detector on-board the Chandra X-ray observatory in very faint mode.}{Caused by interaction of highly energetic particles with the detector, 
the ACIS-I instrumental background is spectrally characterised by 
the superposition of several fluorescence emission lines onto a continuum. To isolate its flux from any sky component, we fitted 
an analytical model of the continuum to observations performed in very faint mode with the detector in the stowed position shielded from the sky, 
and gathered over the eight year period starting in 2001. 
The remaining emission lines were fitted to blank-sky observations of the same period. 
We found $11$ emission lines. Analysing the spatial variation of 
the amplitude, energy and width of these lines has further allowed us to infer that three lines of these are presumably due to an energy correction artefact
produced in the frame store.}
{We provide an analytical model that predicts the instrumental background with a precision of $2\%$ in the continuum and $5\%$ in the lines. 
We use this model to measure the flux of the unresolved cosmic X-ray background in the Chandra deep field south. We obtain a flux of $10.2^{+0.5}_{-0.4} 
\times 10^{-13}\si{\erg \per \square \centi\meter \deg}\si{\per \second}$ for the $[1-2]\si{\kilo\electronvolt}$
band and $(3.8 \pm 0.2) \times 10^{-12}\si{\erg \per \square \centi\meter \deg}\si{\per \second}$ for the $[2-8]\si{\kilo\electronvolt}$ band.}{}
\maketitle

\section{Introduction}\label{introduction}
\begin{figure*}[!ht]
  \begin{center}
   \includegraphics[bb=54 360 558 720,width=11.5cm]{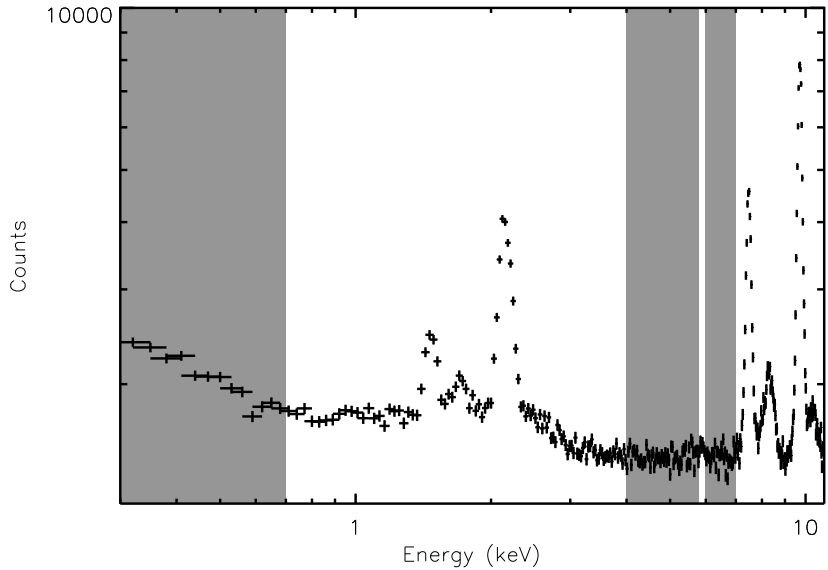}
  \end{center}
 \caption{\textit{VF mode-filtered spectrum of the stowed datasets of periods D plus E. Grey areas 
 identify spectral regions used to model the continuum emission.}}
 \label{fig:presentazione}
 \end{figure*}
 One of the main tasks of observers is the extraction of the target source signal from the data,
 that is separating the source photons from the background photons.
 This requires identifying an adequate procedure to estimate the background.  
 
 For less extended or point-like sources, the background contribution to the observed data can be simply estimated 
 by extracting spectra from  nearby regions that are free from target or other point-source emission. 
 Because the background may have significant 
 spatial variations, the same procedure cannot be applied to extended sources.
 For the latter case, a common approach is to use  ``blank-sky'' datasets.
 These datasets are obtained by stacking the data of a number of observations of relatively empty sky regions with 
 low Galactic emission. While this approach accounts for spatial variation of 
 the background (e.g., vignetting or detector non-uniformity)
 it may not be ideal when high accuracy in the background estimation is required. 
 The total X-ray background flux can be subdivided into three main 
 components that have different spatial variation: the cosmic X-Ray background (CXB), the Galactic local foreground,
 and the particle detector background. Furthermore, while the first two components are vignetted, the last one is expected to be less 
 sensitive, if at all, to the characteristics of the telescope mirrors.
 Thus, an accurate background determination for each specific observation requires the three background components to be estimated individually.
 
 The CXB is the superposition of resolved 
 and unresolved emission coming from distant X-ray sources, such as AGNs 
  (\citealt{giacconi}). Although this component is generated by sources with different redshift and spectral 
distributions, it can be modelled by a power law with a photon index 
$ \Gamma = 1.42$ (e.g., \citealt{gamma_cxb_value}). 

The Galactic local foreground component has been studied using the first sky maps of the soft background obtained 
by ROSAT (\citealt{snowden_rosat}). It can been modelled using two-temperature thermal
emission components (\citealt{backg_compo}) although  
its origin and structure are still under debate.

The instrumental component is related to the instrument itself and  originates from the interaction of high-energy 
particles with the instrument and its electronics. 

For XMM-Newton, \cite{xmm_particle} studied the particle background of the EPIC cameras and proposed an 
analytic model that, in conjunction with the CXB and the Galactic foreground model,
can be used to accurately predict the spatial variation of the background in XMM-Newton observations.

Inspired by their work, we study and characterise the spatial and temporal variations of the very faint (VF) mode Chandra ACIS-I particle background.
Here we develop an analytical model for the Chandra ACIS-I detector background and provide prescriptions on how to combine it with the sky components to 
predict the X-ray background for individual Chandra observations.
We illustrate its application by deriving an accurate estimate of the flux of the unresolved CXB in a Chandra deep field. 

The CCD behaviour found in this work is characteristic only for the front-illuminated CCDs of ACIS-I. We performed a similar 
analysis for the two back-illuminated 
chips of ACIS-S and found that there are significant spectral spatial variations. Thus, a proper characterisation of the 
background of the back-illuminated requires a much higher photon statistic, which is not available yet. 
For this reason we limit the detailed analysis to 
the ACIS-I detector.

 The paper is organised as follows: in Sec. \ref{dataset} we introduce the datasets used in this work, 
 in Sec. \ref{acis_particle_background} we describe the model production methodology and how to apply it to an observation, 
 in Sec. \ref{risu_cdfs} we report results obtained by testing our model on a real observation.
All fits described in this work were performed using maximum-likelihood estimation.
\section{Dataset} \label{dataset}

In this section we briefly describe the datasets used to model the particle background. All the datasets are available at the 
Chandra Background Web-page\footnote{cxc.harvard.edu/contrib/maxim} (CBW). 
Because the Chandra background flux varies with time, the background datasets have been divided into five time periods, A-B-C-D-E (for further 
details see the Markevitch COOKBOOK\footnote{cxc.harvard.edu/contrib/maxim/acisbg/COOKBOOK}, which for convenience we call background-information file,
or BIF). Observations in VF mode allow the most efficient background filtering (for details see \citealt{vikh_webpage}). 
 They are available starting from year $2000$, that is, for periods D and E. Because we are interested in modelling the VF mode background, 
 we used only datasets from these two last periods. Owing to the differences in background flux, focal plane temperature, and observation mode we cannot 
 use datasets from earlier periods.
The Chandra calibration team provides two different background datasets, "stowed" and "blank-sky". 
 
 In stowed configuration, the ACIS detector
is moved to a position that is not exposed to sky, so all the photons detected are expected to come from the particle background. 
Comparison with the Chandra observation of the dark moon confirms that the spectrum of the stowed data is a good representation of 
 the in-focus ACIS particle background (for details see \citealt{mark_cxb}).

The available stowed files have an effective exposure time of $\SI{253}{\kilo\second}$ and $\SI{367}{\kilo\second}$ for periods D and E, respectively. 
From the particle background point of view, CCD0 and CCD1 are similar, except for a difference in the normalisation 
factor\footnote{cxc.cfa.harvard.edu/contrib/maxim/stowed/i01}. Because of the telemetry limits, 
CCD1 has been turned off during the observations in stowed configuration, therefore
the data events of CCD1 reported in the stowed dataset are a randomised copy of the CCD0 events (see \citealt{mark_cxb} for details).

The blank-sky background was obtained by combining selected observations with moderate to low soft Galactic brightness
 and low column absorption $ n_H \sim (1-5) \times 10^{20} \si{\centi\meter}^{-3}$. Point sources were removed from individual observations, 
 and all observation events were combined to obtain a single observation in which the holes left after source exclusion were filled. For further details, 
see the CBW and in particular, the README file on how datasets were assembled. The photons in this dataset are produced by both 
the particle and the X-ray sky background. The total exposure time for periods D and E are 
$t \sim 1.5 \times 10^6 \si{\second}$ and $t \sim 1.55 \times 10^6 \si{\second}$ , respectively. As described in detail below, we used 
the larger statistics provided 
by the D+E blank-sky dataset to refine the background and 
model the spatial variation of the line emission.

To test the accuracy of our background model, we used the Chandra Deep Field South
(CDFS), which represents one of the deepest observation available in the Chandra archive of a sky region without highly extended sources that is removed 
from any Milky Way bright feature. In particular, we used the $43$ observations reported in 
Table \ref{tab:cdfs_data} for a total exposure time of $t \sim 3 \times 10^6$ \textit{s}. 

We reprocessed all CDFS data using CIAO tools, version 
$4.3$ and calibration files version $4.3.3$, following science data analysis threads. The same software package was used to produce 
response files and to clean the data from non-quiescent background periods using \textit{lc\_clean} with a threshold $\sigma = 3$. All datasets were 
filtered by the VF mode status bit. Bad pixels were excluded using the appropriate bad column and 
pixel maps from the CBW. All the spectral fits were performed with XSPEC using the 
appropriate spatially dependent redistribution matrix files (RMF) and ancillary response files (ARF).
The RMFs and ARFs were computed using the CIAO tools in detector coordinate system. 
\section{VF mode ACIS-I particle background}\label{acis_particle_background}
 
To study the ACIS-I particle background, we first analysed the available stowed data for periods D and E. 
As already shown by \cite{mark_cxb}, we confirm that for ACIS-I the spectral shape during these two periods is the 
same and there is only a difference in the normalisation. This result allows us to merge the two datasets to increase the photon statistics.
In Fig. \ref{fig:presentazione} we show the total spectrum extracted from all the CCDs of the stowed data of periods $D+E$. 

 The spectrum presents a continuum with emission lines. 

The continuum emission is clearly visible in the [$0.3-0.7$]~$\si{\kilo\electronvolt}$,  [$4-5.8$]~$\si{\kilo\electronvolt}$ and 
[$6-7$]~$\si{\kilo\electronvolt}$ bands. For convenience, 
from now on, we refer to these bands as the continuum band and to the other energetic regions as line bands.

\begin{figure}
 \begin{center}
\resizebox{\hsize}{!}{\includegraphics[bb=0 0 374 348]{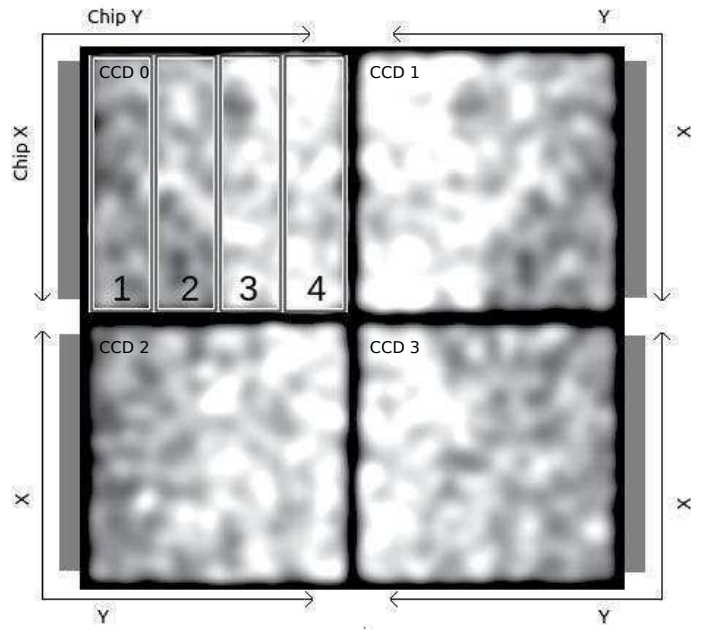}}
 \end{center}
\caption{\textit{ACIS-I photon image of the stowed dataset of periods $D+E$ in the continuum band in detector coordinates. 
Darker and lighter patches indicate lower or higher 
photon number, respectively. The $4$ CCDs are divided by the CCD gap in the middle of the image. The upper-left square 
corresponds to CCD0. The image has been binned with $det x = 4$ and smoothed with a Gaussian kernel with $\sigma=19$ pixels.
Grey external rectangles indicate the position of the frame store areas for each ACIS-I chip. The $x$ and $y$ arrows indicate the chip coordinate reference frames 
for each CCD. The four rectangular areas in CCD0 indicated by increasing number are our analysis strips defined in the text.}}
\label{fig:stowed_grad_ima}
\end{figure}

Using a Gaussian kernel with $\sigma = 19$ {\rm pixels}, we show in Fig. \ref{fig:stowed_grad_ima} a smoothed image
of the stowed background in detector coordinates in the continuum band.
From now on, using the chip coordinate system, we refer to chip-y and chip-x using the labels $y$ and $x$. Both coordinates 
reach from $1$ to $1024$. In Fig. \ref{fig:stowed_grad_ima} we also show the position of the frame store areas
as grey rectangles. To avoid artefacts due to the application of the smoothing kernel to bad pixels and columns, we used 
bad-pixel maps available from the CBW and substituted each bad pixel with a mean 
value computed with the nearest pixels.
The figure indicates that while for each CCD the emission is near-uniform along $x$, 
there is a clear gradient along the $y$ direction. 
To verify and quantify the spectral constancy along $x$ and $y$ we subdivided each CCD into four adjacent 
"analysis" strip regions. Each strip region is a rectangle with ($\Delta x$,$\Delta y$)=($256,1024$) {\rm pixels} to study variations along $x$, or 
($\Delta x$,$\Delta y$)=($1024,256$) {\rm pixels} to study variations along $y$.
We number the strips starting from the edge of the frame stores in case we are studying variations along $y$: strip $1$, $2$, $3$, and $4$ correspond 
to chip coordinate rectangles $(x,y) = (1:1024,1:256)$ {\rm pixels}, $(x,y) = (1:1024,256:512)${\rm pixels}, 
$(x,y) = (1:1024,512:768)$ {\rm pixels} and $(x,y) = (1:1024,768:1024)$ {\rm pixels}, respectively.
To study variations along $x$, we numbered the strips starting from the zero point defined by the chip coordinate system shown in Fig. \ref{fig:stowed_grad_ima}:
strip $1$, $2$, $3$, and $4$ correspond 
to chip coordinate rectangles $(x,y) = (1:256,1:1024)$ {\rm pixels}, $(x,y) = (256:215,1:1024)${\rm pixels}, 
$(x,y) = (512:768,1:1024)$ {\rm pixels} and $(x,y) = (768:1024,1:1024)$ {\rm pixels}, respectively.
In Fig. \ref{fig:stowed_grad_ima} we show the position of the four analysis strips used to study variations along $y$ for CCD$0$. 

\begin{figure*}[!ht]
 \begin{center}
 \includegraphics[bb=0 0 426 307,width=17cm]{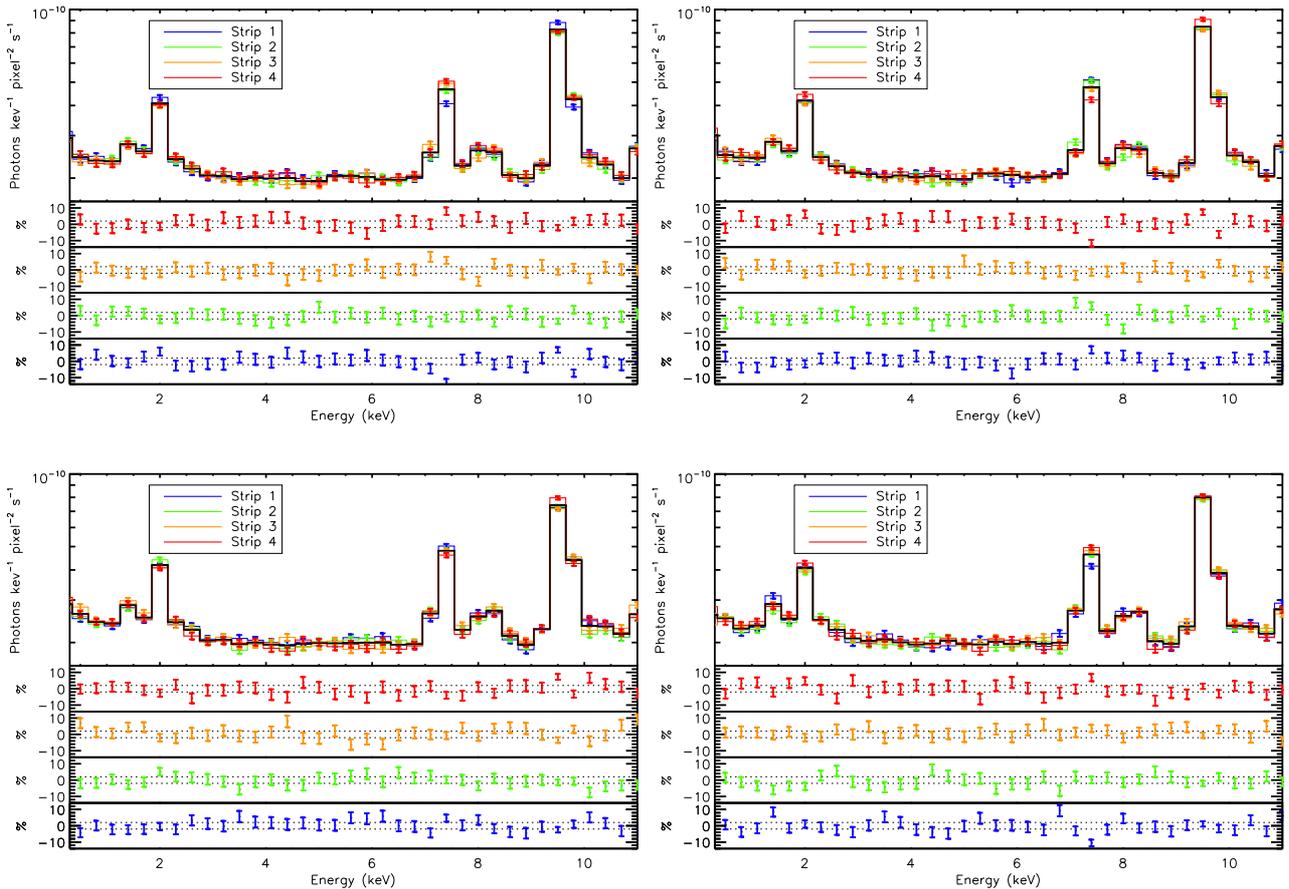}
\end{center}
 \caption{\textit{Upper panel: spectra extracted from the four analysis strips along $x$ plotted as a coloured solid line and total spectra represented as a 
                  black solid line. Lower panels: percentage variation of the strip spectrum with respect to the total spectrum. 
                  Dotted lines are $\pm 2\%$ lines. In 
                  the top-left panel we show the CCD$0$ spectra, in the top-right panel CCD$1$, in the bottom-left and bottom-right panel CCD$2$ and CCD$3$.}}
 \label{fig:alongx0}
\end{figure*}

Each spectrum extracted was normalised by the area, time, and energy.
Then, we compared each strip spectrum with the spectrum extracted from the entire CCD 
to identify variations in the spectral shape with respect to a reference value. We found that spectral variations along $x$ are, within statistical errors, 
lower than $2\%$ in the continuum but in the line bands the variation is greater and is lower than $5\%$.
We show in Fig. \ref{fig:alongx0} the comparison of the four strip spectra with respect to the total chip spectrum for the four chips of ACIS-I. Starting 
from the top-left panel we show CCD$0$-$1$-$2$-$3$, with a bin of 
$\SI{0.3}{\kilo\electronvolt}$. In each upper panel we show the spectra from each strip compared with the total spectra as a black solid line. In 
the lower panels we show the percentage variation, defined as the percentual difference between the spectrum of 
interest and the total one, where the dotted lines denote $\pm 2\%$ levels. 
All the variations for each strip are consistent with the $2\%$ levels. We obtain the same result for the other three chips, as shown in the 
other images of Fig. \ref{fig:alongx0}.

\begin{figure*}[!ht]
 \begin{center}
 \includegraphics[width=17cm,bb=0 0 213 154]{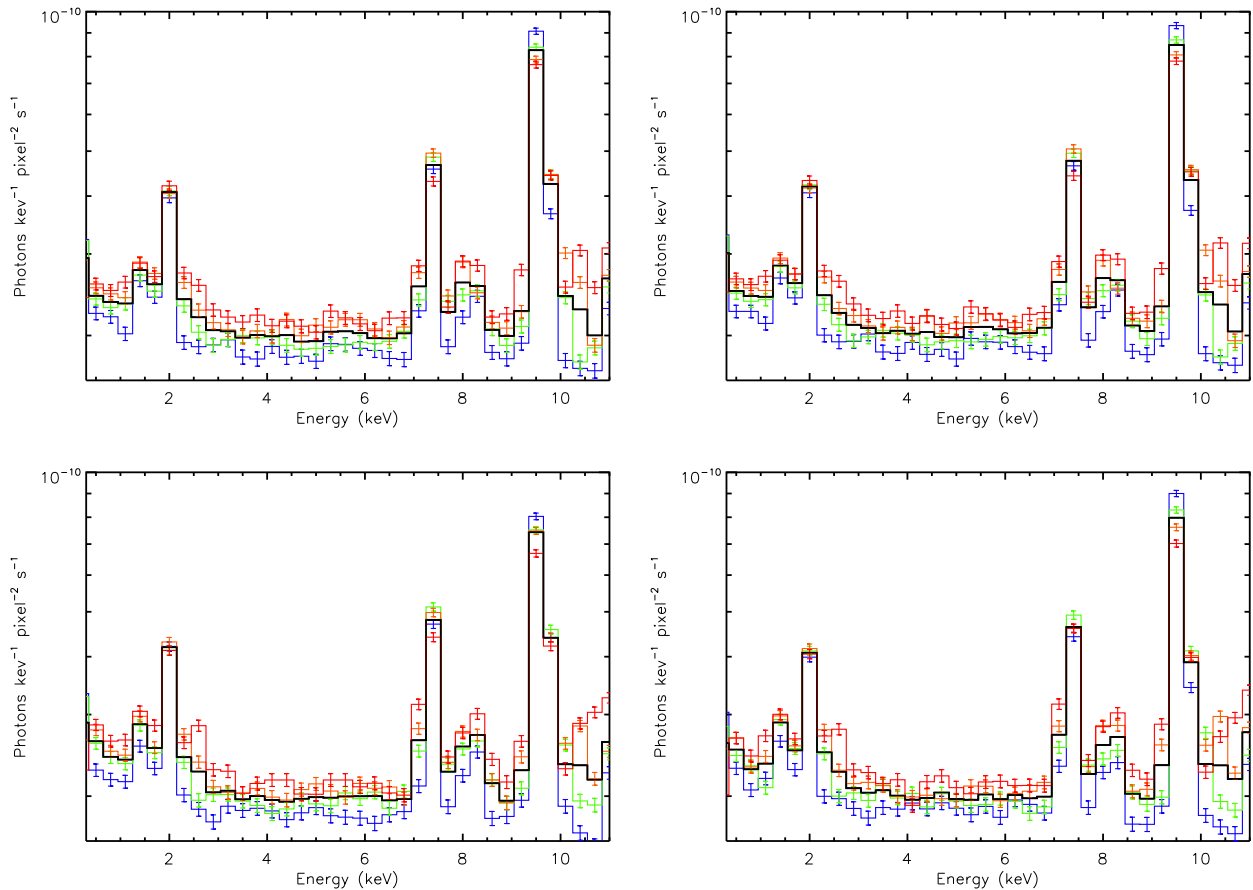}
\end{center}
 \caption{\textit{Same as Fig. \ref{fig:alongx0}, but for the study performed along $y$.}}
 \label{fig:alongy0}
\end{figure*}

We applied the same method to study spectral variations along the $y$ direction for the four chips. As we can see from 
Fig. \ref{fig:alongy0}, where we plot the spectra extracted from the four strips shown in Fig. \ref{fig:stowed_grad_ima} and the total spectrum, 
the spectral shape varies from the lower to the upper strips. In particular, the continuum normalisation increases towards higher $y$ regions, as 
observed in Fig. \ref{fig:stowed_grad_ima},
while lines show a more complex behaviour. We obtain the same results for the other three chips, 
as shown in Fig. \ref{fig:alongy0}, where starting from the top-left panel we show the CCD$0$-$1$-$2$-$3$ spectra.

This observed gradient is most likely due to events collected by ACIS in the frame stores (grey areas in Fig. \ref{fig:stowed_grad_ima}) during the readout phase. 
Even if frame stores are shielded to prevent detection of photons, high-energy particles pass through the coating, and some are registered as X-ray photons. 
Because photons are read starting from low $y$ rows, high $y$ rows are exposed longer to these high-energy particles and, thus, collect more events than low chip $y$ 
rows.

To account for the above findings, we modelled the ACIS-I VF particle background as a function that depends only on the $y$ direction. 
To allow for the possible different coordinate dependence of the line and continuum emission we write the background as a sum of two terms, $C$ and $L$, 
to account for the continuum and line emission, respectively:
\begin{equation}
 F_{i}(E,y) = C_{i}(E,y) + L_{i}(E,y),
\label{eq:generale}
\end{equation}
where $E$ is the photon energy. The particle background may vary from chip to chip, so in our formulas we indicate the CCD ID with the suffix $i$ where 
$i=[0,1,2,3]$. The dimension of $F$ is event per second per CCD pixel. 

As explained below, the particle background was modelled 
with a number of parameters that may or may not depend on the detector position. 
In our formulas, we used the convention of identifying parameters that are spatially constant and depend 
on the position using Roman and Greek letters, respectively.

In the next subsections we describe the model for the continuum and line emission.
\subsection{Model continuum emission}\label{continuum_modelization}

As explained in the previous section, by comparing the spectra from the analysis strips along $y$, we found that they have the same continuum shape 
with different normalisations. 
 This allowed us to model the spectrum in the continuum with a position-independent function multiplied by the position-dependent normalisation factor 
 $\alpha_{i}(y)$, where the suffix $i$ indicates the CCD ID. 
 
We modelled the position-independent term using a power law plus an exponential, so the continuum emission is given by
\begin{equation}
  C_{i}(E,y)=\alpha_{i}(y)  \left( K_1  e^{-A_1  E} + K_2  E^{-A_2} \right),
 \label{eq:continuonuovo}
\end{equation}
where $E$ is energy in $\si{\kilo\electronvolt}$.
To find the parameters $K_1$, $A_1$, $K_2$, $A_2$, we fitted equation \ref{eq:continuonuovo} to the overall spectrum extracted from the sum of the four CCDs 
using the continuum band.  
The result of the fit is shown in table \ref{tab:vari_b}.
\begin{table}[htbp]
 \caption{Best-fit parameters of equation \ref{eq:continuonuovo} to the spectrum of the stowed dataset of periods D+E to the continuum band.}
 \label{tab:vari_b}
 \begin{center}
 \begin{tabular}{cc}
 \hline
 \hline
 Parameter          & Value  \\ 
 \hline
 $K_1$              & $0.1493$ \\             
 $A_1$              & $3.8106\si{\kilo\electronvolt}^{-1}$ \\ 
 $K_2$              & $0.0859$ \\ 
 $A_2$              & $0.0292\si{\kilo\electronvolt}^{-1}$ \\
 \hline
 \end{tabular}
 \end{center}
 \end{table}

\begin{figure}
 \begin{center}
 \resizebox{\hsize}{!}{\includegraphics[bb=54 360 558 720]{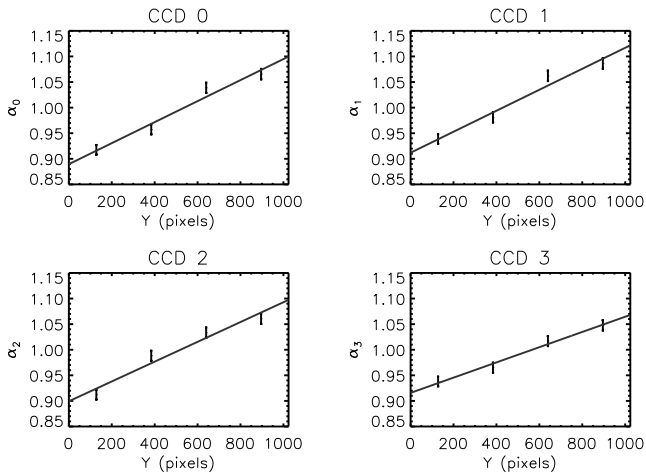}}
 \end{center}
\caption{\textit{Ratio of the mean surface brightness of the analysis strip to the mean surface brightness from the sum of four CCDs. Error bars 
indicate the $68\%$ errors. The four panels correspond to the four CCDs of ACIS-I. Solid lines are the best-fit results of equation \ref{eq:linear_law} 
to the data from table \ref{tab:alpha}.}}
\label{fig:grad_graf}
\end{figure}

To characterise the normalisation parameter $\alpha_i(y)$, for each analysis strip, we calculated the quantity
\begin{equation}\label{eq:alpha}
\alpha = \frac{cnt_{strip}}{cnt_{tot}}\frac{Npix_{tot}}{Npix_{strip}},
\end{equation}
where $cnt_{strip}$ and $cnt_{tot}$ are the total number counts in the continuum band from the strip considered and the sum of four CCDs, respectively, and 
$Npix_{tot}$ and $Npix_{strip}$ are the number of pixels of all four CCDs and in the strip considered, respectively.
In this calculation, we account for all the bad columns and pixels provided in the maps available from the CBW.
The $\alpha$ parameter is a dimensionless quantity that gives the ratio of the mean surface brightness of the analysis strip 
with respect to the mean surface brightness of the entire FOV.
In the ideal case of a spatially uniform background, $\alpha = 1$ is expected in all the analysis strips.
The $\alpha$ values and errors for each strips are reported in Fig. \ref{fig:grad_graf}. 
In this figure, the Y value associated with each strip corresponds to the algebraic mean 
of the $y$ positions of all the pixels used to derive the strip number counts.
Fig. \ref{fig:grad_graf} shows a linear correlation, therefore we fitted the observed $\alpha$ values for each CCD with
\begin{equation}
  \alpha_{i}(y) = \mu_{i}y + \psi_{i}.
  \label{eq:linear_law}
\end{equation}
In table \ref{tab:alpha} we report the best-fit parameters $\mu_{i}$ and $\psi_{i}$, while in Fig. \ref{fig:grad_graf} we plot these best-fit models.

To test the time stability of our models, we split the merged background D+E file into period D and period E.
We repeated the previous analysis for the two datasets independently and found consistent results. 
\begin{table}
 \caption{Best-fit parameters for the $\alpha_{i}(y)$ gradient defined in \ref{eq:linear_law}.}
 \label{tab:alpha}
  \begin{center}
  \resizebox{5cm}{!}{
   \begin{tabular}{ccc}
 \hline
 \hline
    CCD ID ($i$)        &   $\mu$ [$10^{-4}$]  &  $\psi$ [$10^{-1}$] \\
 \hline
       0                &    2.054                           & 8.895   \\
       1                &    2.050                           & 9.119   \\
       2                &    1.936                           & 8.993   \\
       3                &    1.488                           & 9.157   \\
  \hline
   \end{tabular}
   }
  \end{center}
 \end{table}
 \subsection{Model of the emission lines}\label{lines_modelization}

 The spectrum in Fig. \ref{fig:presentazione} shows at least $11$ emission lines in the 
$[0.3-11]\si{\kilo\electronvolt}$ band. In this paragraph we describe how we modelled these lines and their spatial behaviour. 
Similarly to the continuum, we extracted the spectra for each CCD using the analysis strips defined above, and studied the line 
behaviour. 
We found that five lines, three in the $[1-2]\si{\kilo\electronvolt}$ band and two at 
$\sim \SI{5}{\kilo\electronvolt}$ and $\sim \SI{8.3}{\kilo\electronvolt}$ are position-independent, while the other six appear to change 
both in energy and normalisation. 
A more detailed inspection of these three position-dependent lines revealed that they can be split into prominent major and minor components. 

To properly characterise the line behaviour, we used the blank-sky dataset, because its effective exposure is $\sim$four times higher than the stowed background. 
The increased photon number improves the statistics to study the line spatial variations.
The drawback is that compared with the stowed background, it also contains the sky background emission. 
To use this dataset, we therefore also modelled the sky components that need to be convolved with the mirror response.
In addition to the sky background continuum emission, there are strong line emissions at lower 
energies ($\sim 0.5 \si{\kilo\electronvolt}$ e.g. \citealt{galactic_emission}), which do not blend with the instrumental lines 
analysed and do not affect our analysis.

We modelled the sky background components using the results from \cite{backg_compo} to account for the Galactic emission, 
and results from \cite{gamma_cxb_value} to account for the CXB. 
In particular, we used an absorbed power law with photon index $ \Gamma =1.42$ for the CXB 
while for the emission from beyond the Galactic absorption we used absorbed thermal APEC models with temperatures of 
$kT = \SI{0.14}{\kilo\electronvolt}$ and $kT=\SI{0.248}{\kilo\electronvolt}$,
respectively.
In the two thermal models, we fixed the redshift to $z=0$ and the metallicity to the solar value. 
The absorption column density $n_{H}$ was obtained by fitting the spectra extracted from the VF blank-sky. We found 
$n_{H} = \SI{1.63d20}{\centi\meter}^{-3}$, which agrees with the value range [$(1-5) \times 10^{20} \si{\centi\meter}^{-3}$] reported in the BIF. 
We normalised the components by fitting the models to the spectrum extracted from the sum of the four CCDs of the blank-sky dataset in the range $[0.5-11]\si{\kilo\electronvolt}$. 
 \subsubsection{Constant lines}
 To model the five position-independent lines, we extracted the spectra from the entire FOV of the blank-sky, fixed the continuum using the 
 continuum model described in Sec. \ref{continuum_modelization}, and fitted the lines with five Gaussian:
\begin{equation}
D(E) = \sum^{5}_{n=1} A_{n}e^{- \frac{(E-E_{n})^{2}}{2  S_{n}^{2}} },
\label{eq:fixed_lines}
\end{equation}
where  $A_n, S_n$, and $E_n$ are the free parameters that correspond to amplitude, width, and position of the lines, respectively.
The results of the fit are reported in table \ref{tab:righe}.
\begin{table}[htbp]
 \caption{Best-fit values of the five constant line parameters (equation \ref{eq:fixed_lines}). Units of amplitude $A_n$ are photon per CCD pixel per second.}
 \label{tab:righe}
 \begin{center}
 \begin{tabular}{cccc}
 \hline
 \hline
  Line index   & $E_{n}$ $[\si{\kilo\electronvolt}]$ & $\sigma_{n}$ [$\si{\kilo\electronvolt}$ $10^{-1}$] & $A_n$ [$10^{-2}$]   \\ 
 \hline
 $1$           & $1.124$                 & $1.359$                             & $0.410$ \\
 $2$           & $1.486$                 & $0.492$                             & $0.783$ \\ 
 $3$           & $1.814$                 & $1.323$                             & $0.601$ \\ 
 $4$           & $5.900$                 & $0.331$                             & $0.103$ \\
 $5$           & $8.305$                 & $1.919$                             & $1.459$ \\ 
 \hline
 \end{tabular}
 \end{center}
 \end{table}
 
As shown in Sec. \ref{varying lines}, the line at $\SI{8.305}{\kilo\electronvolt}$ 
may be contaminated in the high $y$ region by a spatially variable line. To minimise the contamination of the
$\SI{8.305}{\kilo\electronvolt}$, we therefore only fitted the lowest $y$ regions for this line (i.e. analysis strip $1$).
 \subsubsection{Position-dependent lines}\label{varying lines}
 \begin{figure}[!ht]
 \begin{center}
 \resizebox{\hsize}{!}{\includegraphics[bb=54 360 300 720]{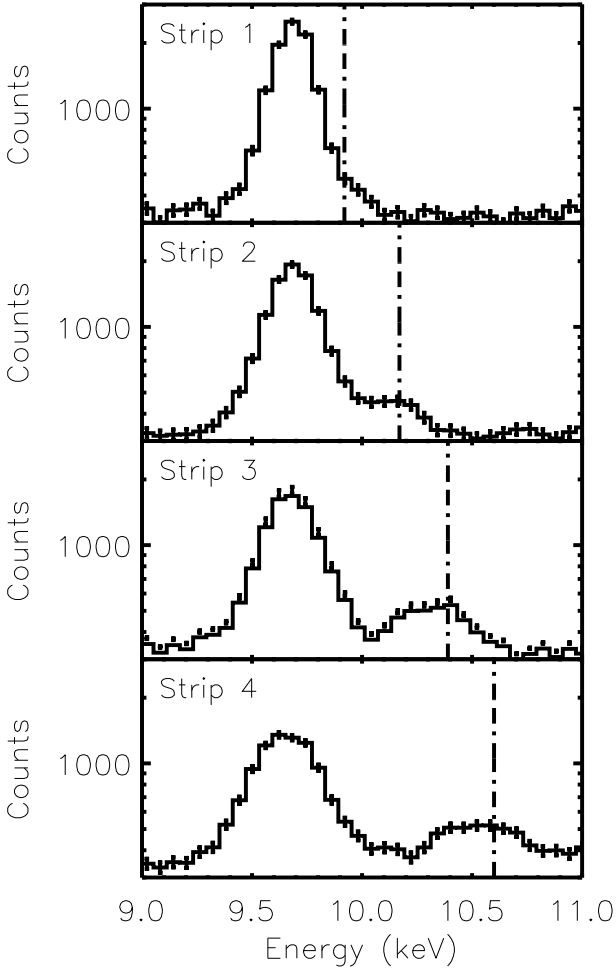}}
\end{center}
\caption{\textit{Spectra extracted from the blank-sky dataset of CCD0 in the four analysis strips. From top to bottom we show the spectra 
from analysis strip 1 to analysis strip 4. The vertical dotted line is the predicted energy position; for details see Sec. \ref{varying lines}.}}
\label{fig:esempio_riga_alta}
\end{figure}
The six position-dependent lines arise from three lines that split into mother-daughter systems.
The mother lines are approximately located at $\SI{2.1}{\kilo\electronvolt}$ (Au, Mg), $\SI{7.5}{\kilo\electronvolt}$ (Ni, $K_{\alpha}$) and 
$\SI{9.7}{\kilo\electronvolt}$ (Au), while the daughter lines have a slightly higher energy than their mothers.
The mother-daughter line split is illustrated in Fig. \ref{fig:esempio_riga_alta}. In this figure 
we compare the spectra in the $[9-11] \si{\kilo\electronvolt}$ band obtained from the four analysis strips of CCD0.

The figure shows that while there is only one line (the mother line 
at $\sim \SI{9.7}{\kilo\electronvolt}$) in the low-$y$ spectrum, at higher $y$, some of the mother line photons are shifted up to higher energies 
and form a daughter line.
Fig. \ref{fig:esempio_riga_alta} clearly shows that the median energy of the daughter line (vertical dashed-dotted line) 
is higher away from the frame store edge (i.e. high $y$).

The mother-daughter line split is an artefact of the Charge Transfer Inefficiency (CTI) correction. 
During the read-out procedures, a certain number of photo-electrons are lost because of CTI. 
As photon energy is a function of the number of photo-electrons or pulse height amplitude (PHA), this effect causes the detected photon 
energy $E_{raw}$ to be lower than the true energy. To restore the true PHA, events were reprocessed \textit{a posteriori} to correct for the CTI effect (for 
more details see, e.g., \citealt{cti_esempio}).
The mother-daughter line effect arises because these background lines
form both in the image area and in the frame store area of the chip. Being shielded, the latter has not been damaged in the way of the image area 
by high-energetic protons. Unfortunately, photons detected in the two areas are indistinguishable, therefore we applied
the same CTI correction as for the imaging area to all the photons detected in both areas. As a consequence, the photons 
registered in the frame store are incorrectly shifted to higher energies, producing the artefact that we observe as the daughter line. 

\begin{figure}[!ht]
 \begin{center}
 \resizebox{\hsize}{!}{\includegraphics[bb=54 360 558 720]{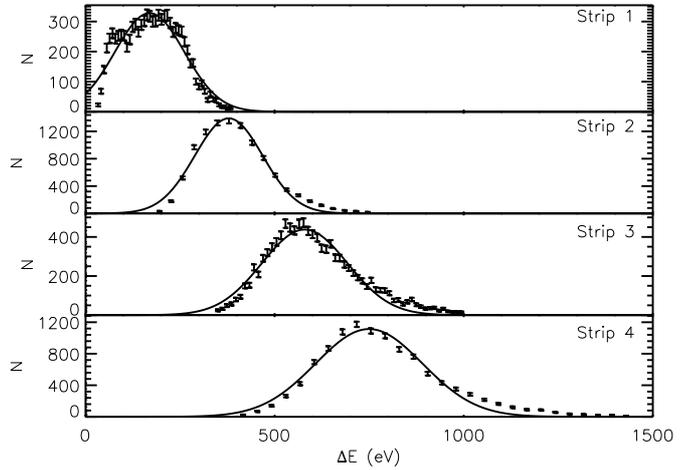}}
\caption{\textit{Displacement energy distribution induced by CTI correction to the photons in the MDSB 2 ($7.3-8.4 keV$), CCD$0$. Solid black line is is the gaussian fit 
 to the data ($\Delta E$).}}
 \label{fig:example_pha}
 \end{center}
\end{figure}

The energy shift depends on the $y$ position of the detected photon. It increases with $y$, since CTI increases with 
distance from the frame store edge. While the CTI correction in the ACIS telemetry processing software is implemented on an individual chip column basis, 
an average correction would be adequate.
We retrieved the average energy displacement distribution induced by the CTI correction using the 
pulse height before ($pha_{raw}$) and after the CTI 
correction ($pha_{crt}$).
To do this we identify three narrow energy bands containing the respective mother-daughter line systems. 
The three mother-daughter system bands (MDSB) used are MDSB1 = $[2-3]\si{\kilo\electronvolt}$, MDSB2 = $[7.3-8.4]\si{\kilo\electronvolt}$, and 
MDSB3 = $[9.5,10.5]\si{\kilo\electronvolt}$. For each MDSB, we extracted the spectra in different analysis strips and determined the 
distribution of the photon energy displacement due to CTI correction by computing the histogram of the following quantity:
\begin{equation}\label{eq:phi_displacement}
 \Delta E = \frac{pha_{crt} - pha_{raw}}{pha_{raw}} E.
\end{equation}
For illustration purposes, in Fig. \ref{fig:example_pha} we give an example of these histograms by showing
the energy displacement distribution in the MDSB2 for the four analysis strips of CCD$0$. 
\begin{figure*}[!ht]
 \begin{center}
  \includegraphics[bb=0 0 375 267,width=16cm]{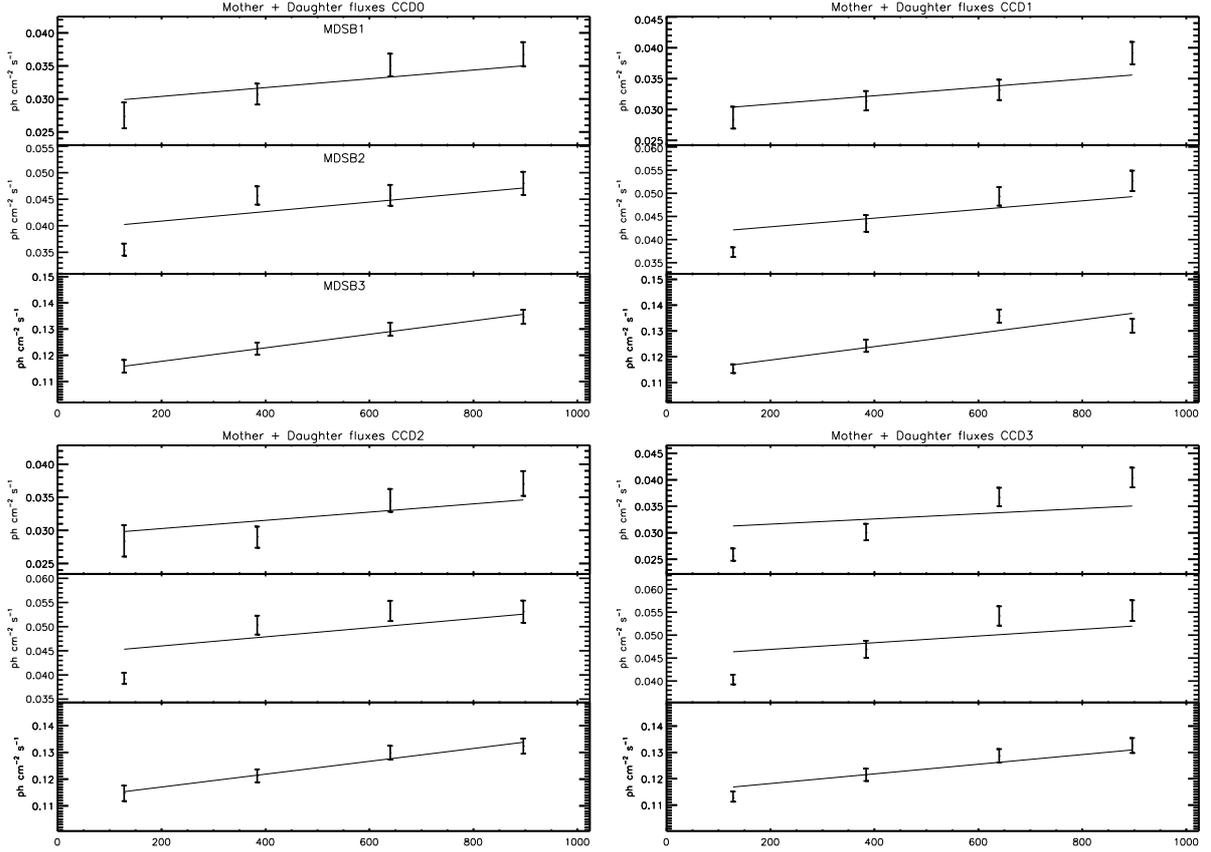}
 \end{center}
 \caption{\textit{Sum of the fluxes of mother and daughter lines for each MDSB system as a function of $y$ position. Overplotted are the slopes of the continuum 
 gradient reported in table \ref{tab:alpha}. Each gradient slope has been multiplied by the median value to match the data.}}
 \label{fig:amplitudes_y}
\end{figure*}

As expected, the median 
energy displacement increases from strip $1$ (lowest $y$) to strip $4$ (highest $y$). 
This shows that the photons 
detected in the frame store that do not require CTI correction are erroneously displaced to higher energies: in each analysis strip spectrum, 
the width of the daughter line is equal to the relative width of the $\Delta E$ distribution, while the 
median daughter-line energy position will be equal to the sum of the mother-line energy and the median of the energy displacement distribution.
To be able to predict the width and position of the daughter line as a function of $y$ for each CCD, we modelled these energy displacement distributions 
in each of the three narrow 
energy intervals. 
We indicate the mother-daughter line system with subscript $l$ ($l=1,2,3$). 
For each analysis strip, we approximated 
the $\Delta E$ distribution with a Gaussian that we fitted to the $\Delta E$ histogram to derive the width $\nu_{i,l}$ and the median 
displacement energy $\delta_{i,l}$. As an example, in Fig. \ref{fig:example_pha} we overlay the Gaussians resulting from the fitting procedure. 
Then, we took the spectrum from each analysis strip and, using XSPEC, fitted the mother-daughter line system in each 
MSDB using two Gaussians:
\begin{equation}
 S_{i,l}(E) = \Phi_{i,l} e^{\frac{(E-B_{i,l})^{2}}{2  Q_{i,l}^{2}}} +  \Theta(E-B_{i,l}) \phi_{i,l} e^{\frac{[E-(B_{i,l}+\delta_{i,l})]^{2}}{2 \nu_{i,l}^{2}}},
\label{eq:mother_and_daughter}
\end{equation}
where the first and second Gaussian correspond to the mother and daughter line, respectively. Because it is 
produced by moving up of photons of the mother line, the daughter line cannot contain photons with $E<B_{i,l}$. 
To account for this, we multiplied the Gaussian of the daughter-line with the Heaviside function $\Theta(E,B_{i,l})$ 
(see second term of equation \ref{eq:mother_and_daughter}).

We fitted equation \ref{eq:mother_and_daughter} leaving the amplitude of mother, $\Phi_{i,l}$, and daughter line $\phi_{i,l}$, 
the energy position, $B_{i,l}$, and width, $Q_{i,l}$, of the mother lines as free parameters. 
Consistently with the above interpretation of the mother-daughter artefact, we found that $B_{i,l}$ and $Q_{i,l}$ remain constant along the analysis strip, 
while the other fitted parameters vary.
We investigated the spatial behaviour of the sum of mother-line and daughter-line fluxes and found that the total flux along $y$ 
for each MDSB system follows the same continuum gradient as shown in Sec 
\ref{continuum_modelization}.  As we can see from Fig. \ref{fig:amplitudes_y}, where we show the sum of the fluxes as a function of $y$ and overplotted the gradient slopes 
reported in table \ref{tab:alpha}, the slope reproduces the spatial behaviour of the total flux within the statistical uncertainties.

\begin{figure*}[!ht]
 \begin{center}
 \includegraphics[bb=54 360 558 720,width=13.1cm]{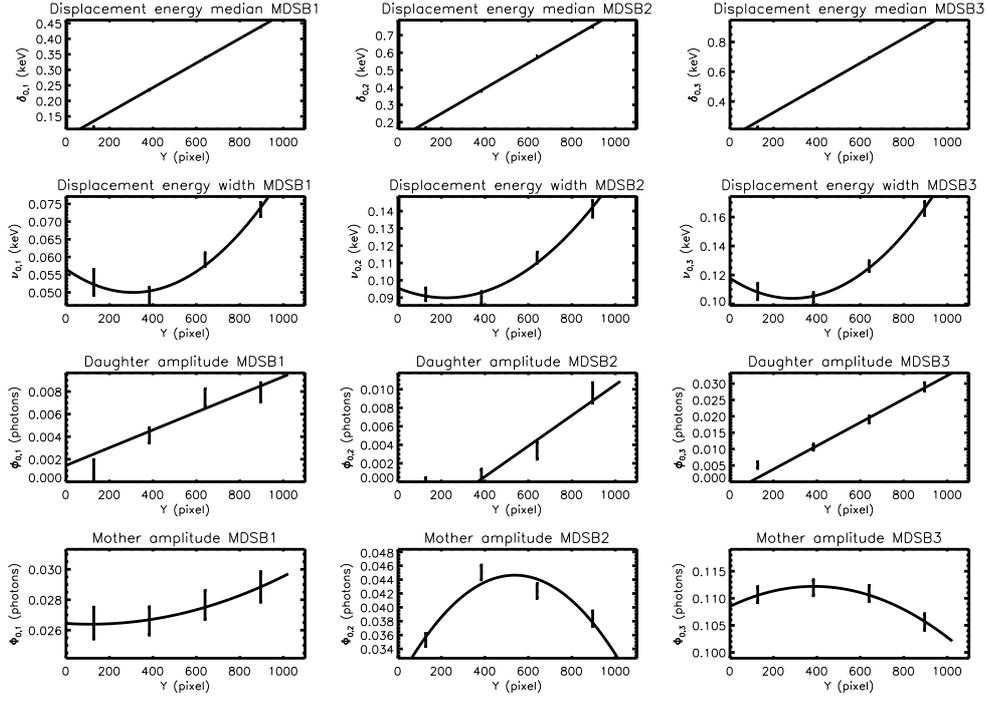}
 \caption{\textit{Position-dependent quantities of the MDSBs for CCD0. Rows are counted from the top. First row: median energy displacement of 
 the daughter line ($\delta$). Second row: width of the daughter line ($\nu$). Third row: amplitude of the mother line ($\Phi$). 
 Fourth row: amplitude of the daughter line ($\phi$). The columns from left to right correspond to the first, seconds and third mother-daughter 
 system, respectively. Continuous lines are the best-fit model to the data points.}}
 \label{fig:graf_risultati_0}
 \end{center}
\end{figure*}

\begin{figure*}[!ht]
 \begin{center}
 \includegraphics[bb=54 360 558 720,width=13.1cm]{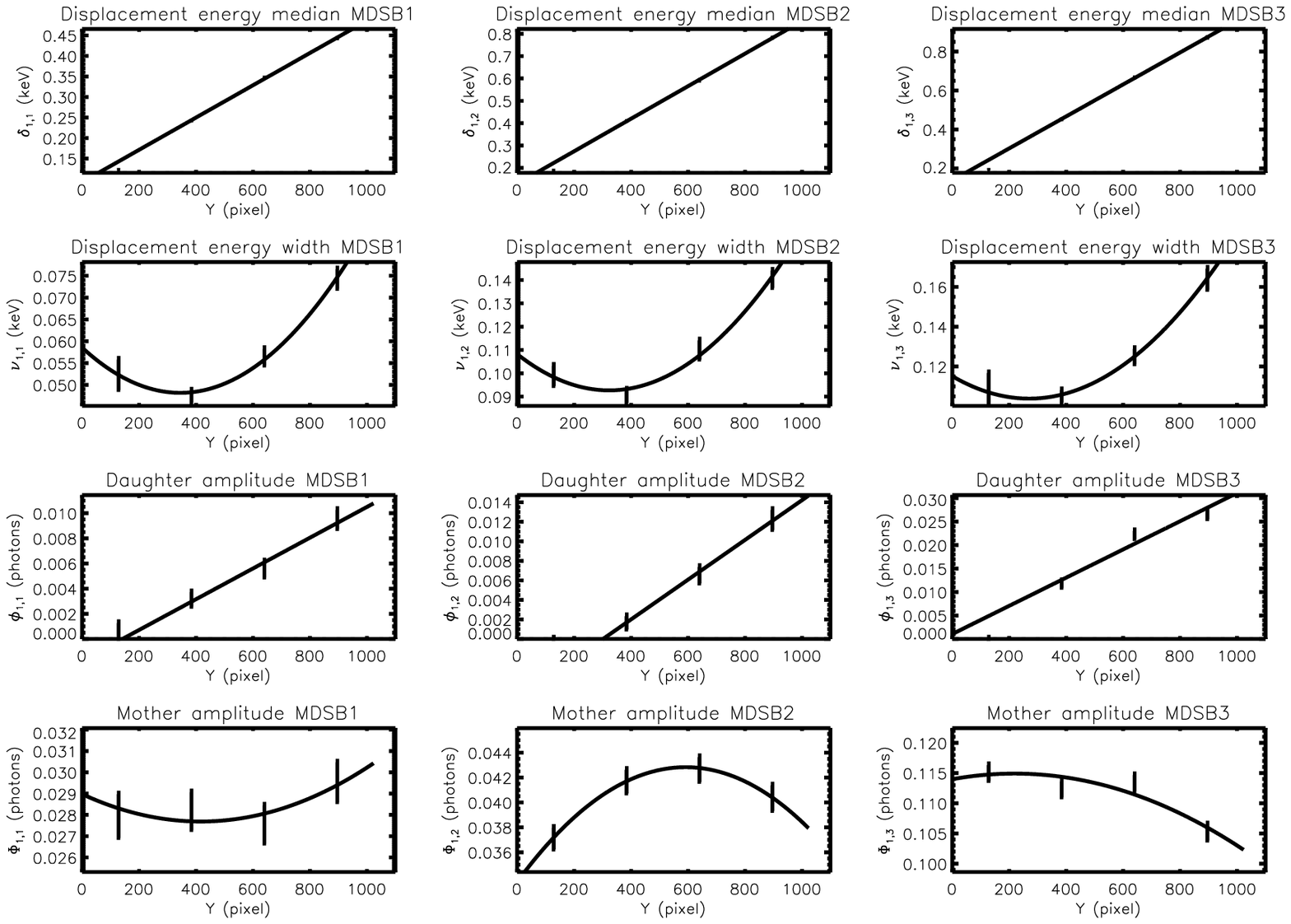}
 \caption{\textit{Same as Fig. \ref{fig:graf_risultati_0}, but for CCD $1$.}}
 \label{fig:graf_risultati_1}
 \end{center}
\end{figure*}

\begin{figure*}[!ht]
 \begin{center}
 \includegraphics[bb=54 360 558 720,width=13.1cm]{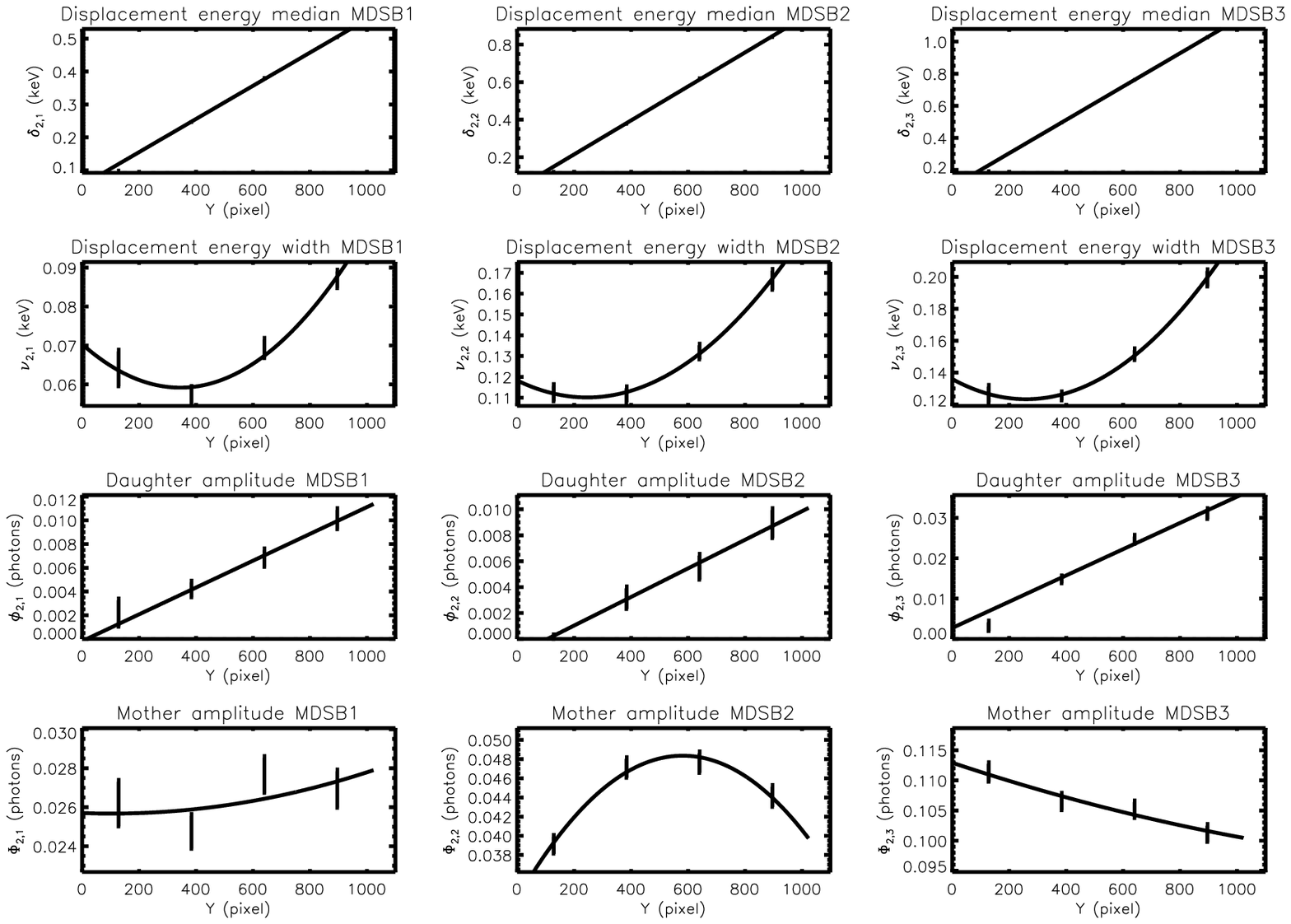}
 \caption{\textit{Same as Fig. \ref{fig:graf_risultati_0}, but for CCD $2$.}}
 \label{fig:graf_risultati_2}
 \end{center}
\end{figure*}

\begin{figure*}[!ht]
 \begin{center}
 \includegraphics[bb=54 360 558 720,width=13.1cm]{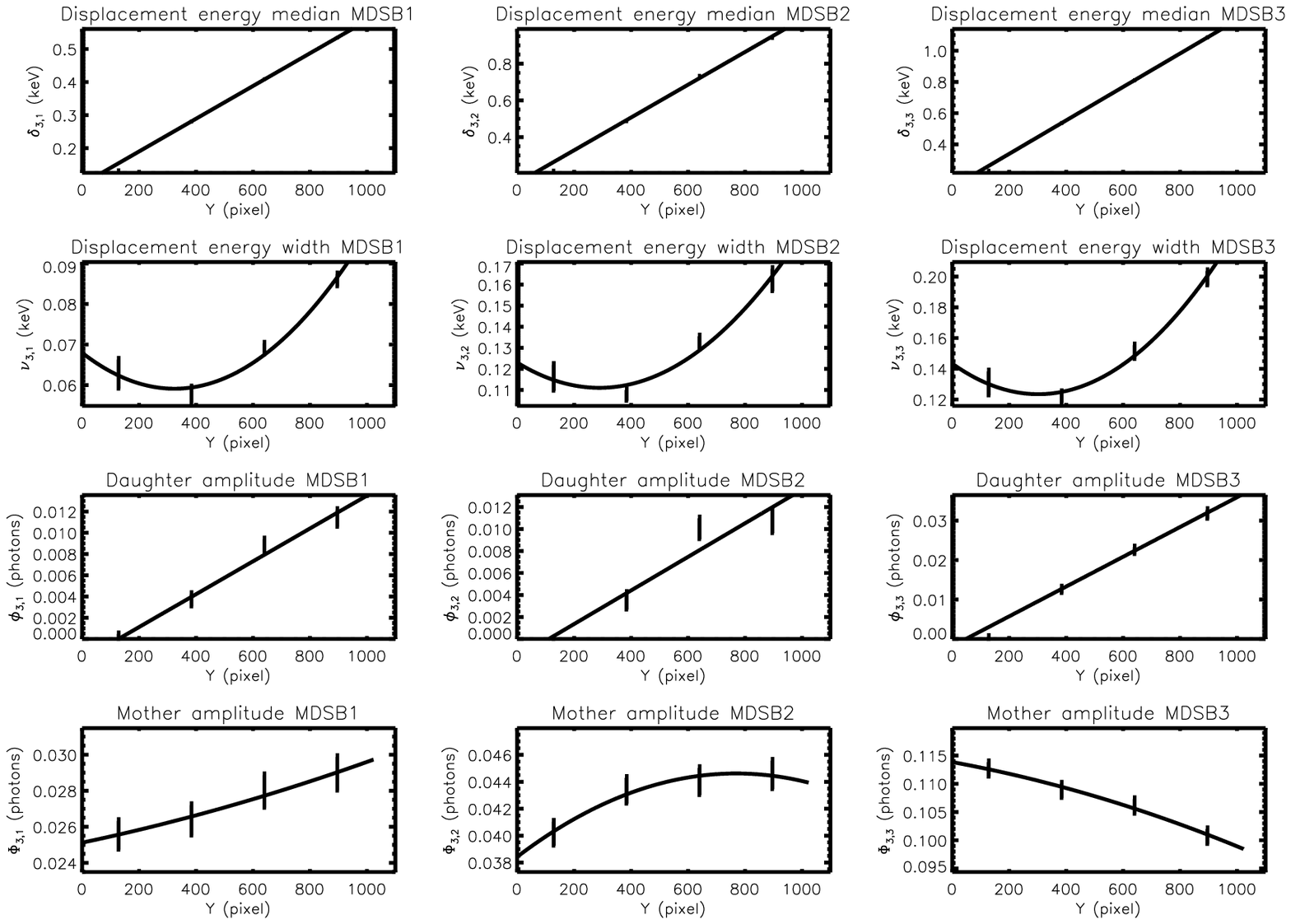}
 \caption{\textit{Same as Fig. \ref{fig:graf_risultati_0}, but for CCD $3$.}}
 \label{fig:graf_risultati_3}
 \end{center}
\end{figure*}

In figures \ref{fig:graf_risultati_0}-\ref{fig:graf_risultati_3}, we report the best-fit values
with their $68\%$ uncertainties of the four varying parameters. Figure \ref{fig:graf_risultati_0} to 
\ref{fig:graf_risultati_3} correspond to CCD ID$=0$ to $3$, respectively. 
The four varying parameters are reported in the four rows, and the columns correspond to the three different mother-daughter line systems considered.
Figures \ref{fig:graf_risultati_0} to \ref{fig:graf_risultati_3} show a linear dependence on $y$ for two parameters (energy displacement $\delta_{i,l}$ 
and daughter amplitude $\phi_{i,l}$) and a quadratic dependence for the other two (the daughter width $\nu_{i,l}$ and the mother amplitude $\Phi_{i,l}$). 
So we modelled the $y$ dependence using the following quadratic function:
\begin{equation}
  \Gamma_{i,l}(y)=\gamma^{2}_{i,l}  y^2 + \gamma^{1}_{i,l}  y  + \gamma^{0}_{i,l},
 \label{eq:parameter2}
\end{equation}
where $\Gamma_{i,l}$ is the particular quantity modelled (i.e. $\delta_{i,l}$ or $B_{i,l}$). 
The fits were made by leaving $\gamma^{2}_{i,l}$,  $\gamma^{1}_{i,l}$, and  $\gamma^{0}_{i,l}$ as free parameters for the relations that are quadratic, 
setting $\gamma^{2}_{i,l} = 0$ for the relations that are linear, and setting to $0$ $\gamma^{2}_{i,l}$ and $\gamma^{1}_{i,l}$ 
for constant relations. 
Results of the fits are reported in tables \ref{tab:ccd0_par}-\ref{tab:ccd3_par} and overlaid on the data points 
in Figs. \ref{fig:graf_risultati_0} to \ref{fig:graf_risultati_3}.

The line emission term, $L(E)$ in equation \eqref{eq:generale}, is then written as the sum of the constant plus the mother-daughter line term:
\begin{equation}
 L_{i}(E,y) = D(E) + \sum_{l=1}^{3}S_{i,l}(E,y).
\end{equation}

\subsection{Normalisation and weighting prescription}\label{prescription}

Even if the spectral shape of the particle background does not vary, its normalisation is a strong function of the observation time. 
This means that if it is to be used for the analysis of a specific observation, our particle background 
model needs to be renormalised. The choice of the energy bound within which this normalisation is performed 
is a compromise between (1) minimising any variation of the background spectral shape across the FOV, (2) minimising the flux
of non-background components, and (3) maximising the statistics of the dataset. Here we used the high-energy end of the spectrum, 
consistently with the standard prescription for normalising the blank-field background, $> \SI{9.5}{\kilo\electronvolt}$, 
since at these energies the ACIS-I effective area decreases 
by a factor of $\sim 100$ (see \textit{The Chandra proposer's observatory guide} for more details). 
The choice of the upper band is delicate. As shown in Fig. \ref{fig:presentazione}, this part of the spectrum contains a number of strong lines, 
for which small energy shifts may result in strong variations for the normalisation. To minimise the effect related to prediction errors 
for the mother-daughter line system at $E > \SI{9}{\kilo\electronvolt}$, we used an iterative approach. We extracted the photons 
from the whole FOV of the blank-sky dataset in several bands, [$9.5 - (9.5+0.1*k)$] $\si{\kilo\electronvolt}$ with $k=1,2...15$, and compared them 
with the prediction of our model to find the one that shows the smallest difference. 
We found that the best band that should be used to normalise the background is [$9.5-10.6$] $\si{\kilo\electronvolt}$.
\begin{figure*}[!ht]
 \begin{center}
\includegraphics[bb=54 360 558 720,width=17cm]{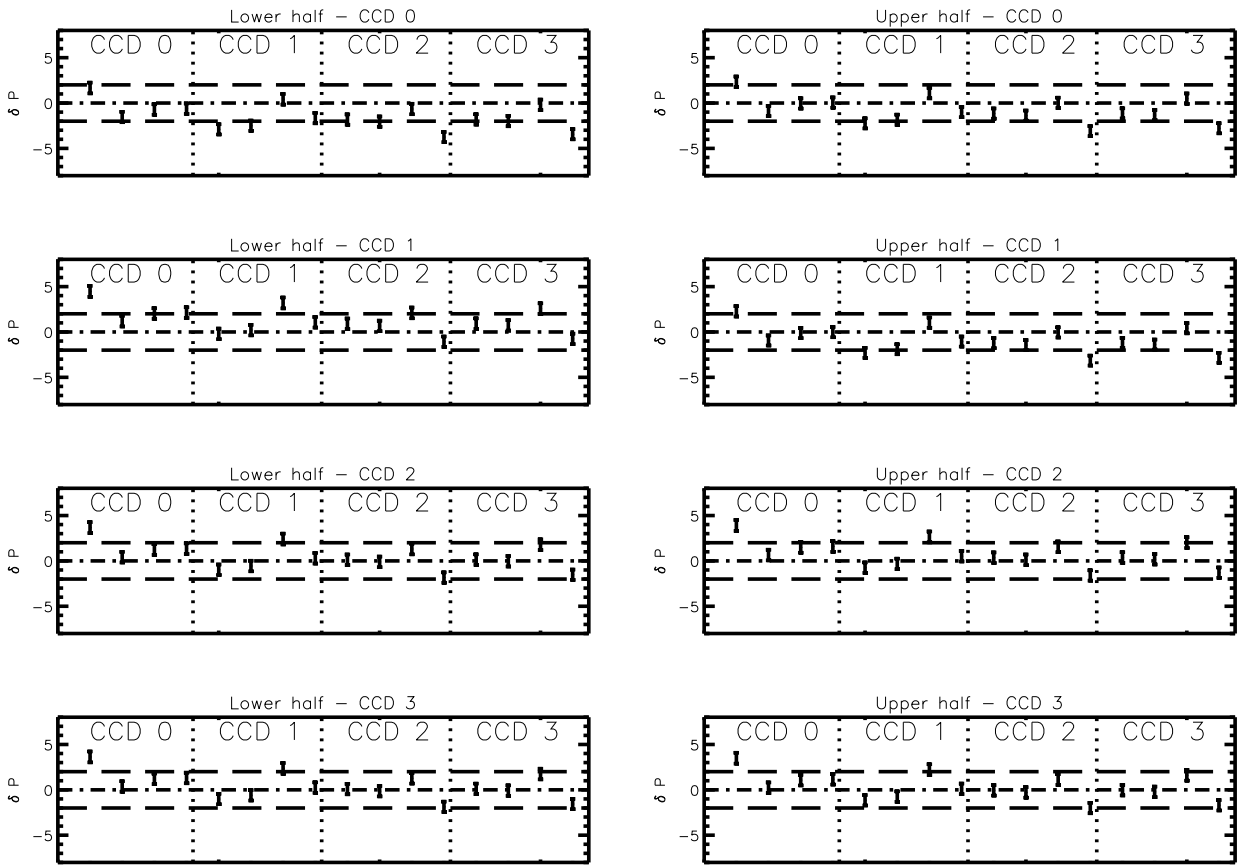}
\end{center}
\caption{\textit{Test of the stability of the background model as a function of the detector region used to normalise the background.
$\delta P$ is the percentile variation of the observed blank-sky background with respect to the background model in the normalisation band ($9.5-10.6 keV$). Points correspond 
to the different analysis strips. The eight panels correspond to the eight different regions used to normalise the background (see panel title). Dashed lines 
indicate the accuracy of $\pm 2\%$.}}
\label{fig:test_blank_sky}
\end{figure*}

To normalise the particle background, one should use the entire field after removing the bright
sources to avoid contamination from event pileup. 
Depending on the position and size of the target, this may
require using a different detector area to normalise the background. Thus, we tested the stability of the proposed model 
using the photons from different detector areas to perform the normalisation. To do that, we took the blank-sky dataset and normalised our 
background model using the lower half of CCD 0 
$(x,y) = (1:1024,1:512)$ (that is, using $1/8$ of the full detector area to normalise the model for the whole detector). Then, 
for each analysis strip, we calculated the percentile variation $\Delta P$ in the $[9.5,10.6] \si{\kilo\electronvolt}$ 
of the model normalised in this way with respect to the data: $\Delta P = \frac{Data-Model}{Model}$.
The result is shown in the upper-left corner of Fig. \ref{fig:test_blank_sky}.

We repeated the test by normalising the background using the upper region of CCD 0 $(x,y) = (1:1024,512:1024)$ and for the lower 
and upper regions of CCD ID $1,2$, and $3$. The results are reported in Fig. \ref{fig:test_blank_sky}.
The figure illustrates that regardless of the actual region used to normalise the background, our model predicts 
the background level in the normalisation band with an accuracy better than $2\%$.

In the analysis of real observations, it is highly likely that regions larger than the one used 
for this test are selected. 
\begin{figure*}[!ht]
 \begin{center}
 \includegraphics[bb=0 -1 540 474,width=12cm]{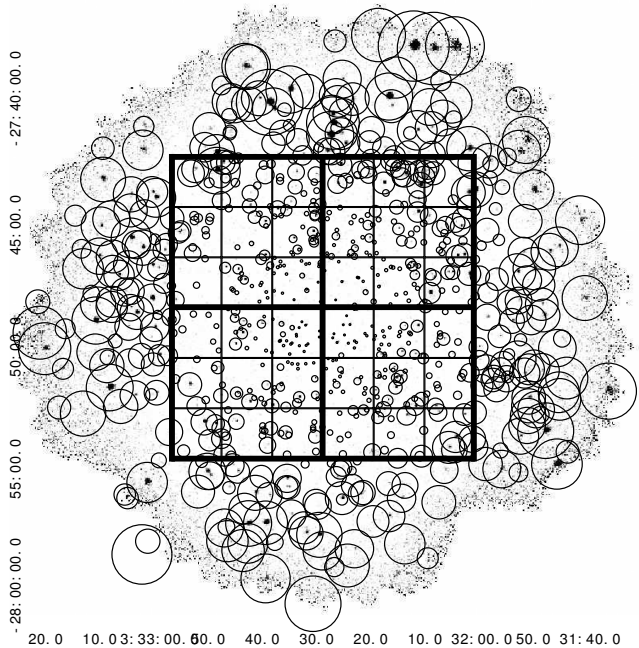}
\end{center}
\caption{\textit{Image of the Chandra deep field south in the $[0.5-2.5]keV$. Circles are the exclusion regions used to remove 
point sources. The black square indicates the regions used for the test of particle background accuracy shown in Figs. \ref{fig:ratio1} to 
\ref{fig:ratio3}. The bottom-left square corresponds to the bottom left square in Figs. \ref{fig:ratio1} to \ref{fig:ratio3}.}}
\label{fig:mappa_cdfs}
\end{figure*}
\section{Analysis of the Chandra deep field south}\label{risu_cdfs}

In this section we use the Chandra Deep Field South (CDFS) observations to quantify the accuracy of our background model. 
The CDFS has a low $N_{H}$ and is located away from any Milky Way bright features.
We retrieved and merged all the ACIS-I VF observations available in the Chandra data archive. The list 
of the observations used is reported in table \ref{tab:cdfs_data}. The total exposure time is $\sim \SI{3.4d6}{\second}$.

We removed contribution from point sources using a mask obtained with the catalogue of $740$ identified X sources of \cite{tozziland}. 
To remove source photons coming from the point spread function (PSF) wings, we used the conservative approach of \cite{mark_cxb}. 
It consists of removing for each point source a circular region whose radius depends both on the source flux $f$ and the PSF broadening. 
We set the radius to be equal to a factor times the $90\%$ encircled-energy radius that depends on the off-axis angle as follows: 
\begin{equation}
 r_{90} = 1'' + 10'' \left( \frac{\theta}{10'} \right)^2,
\label{eq:psf}
\end{equation}
where $\theta$ is the distance from the optical axis \cite{mark_cxb}. As multiplication factor, we used 
$2$, $4.5$, $6$, and $9$ for sources with total photon flux $f$ in the $[0.5-8] \si{\kilo\electronvolt}$ band of  
$f<0.02 \times 10^{-3} \si{\count \per \second}$, $0.02 \times 10^{-3} \si{\count \per \second} \le f <0.2 \times 10^{-3} \si{\count \per \second}$, 
$0.2 \times 10^{-3} \si{\count \per \second} \le f < 2 \times 10^{-3} \si{\count \per \second}$, and 
$ f \ge 2 \times 10^{-3} \si{\count \per \second}$, respectively.
In Fig. \ref{fig:mappa_cdfs}, we overlay the exclusion regions (due to point source removal) on the $[0.5-11]\si{\kilo\electronvolt}$ image of the CDFS.

To normalise the background model, we tried to mimic a procedure that can be applied to the observation of a generic extended target, which is 
usually located near the centre of the FOV. For this reason, we chose to normalise the background in an annulus centred on the centre of the FOV for 
the combined image ( (J2000) $RA=03:32:28.37$ $DEC=-27:48:10.1$) and $\SI{5}{\arcmin} \le r \le \SI{16}{\arcmin}$. 
The outer radius was chosen to be larger than the FOV to ensure that we used all the available photons.

We created the model for the particle background as described above and used it without the application of ARF.
We then added to the normalised model for the particle background the sky component model described in Sec. \ref{lines_modelization} with 
free normalisation for APEC and 
power-law components and fitted it to 
the spectrum extracted in the same annulus. We fixed the absorption column density to the Galactic value, that is corresponding to the CDFS direction, 
$N_{H}=8.8 \times 10^{19}\si{cm}^{-3}$ (see \citealt{cdfs_nh}). This allowed us to completely characterise the spatial variation of the total background 
(particle plus sky components) over the whole FOV.

\begin{figure*}[!ht]
 \begin{center}
  \includegraphics[bb=0 0 565 406,width=15cm]{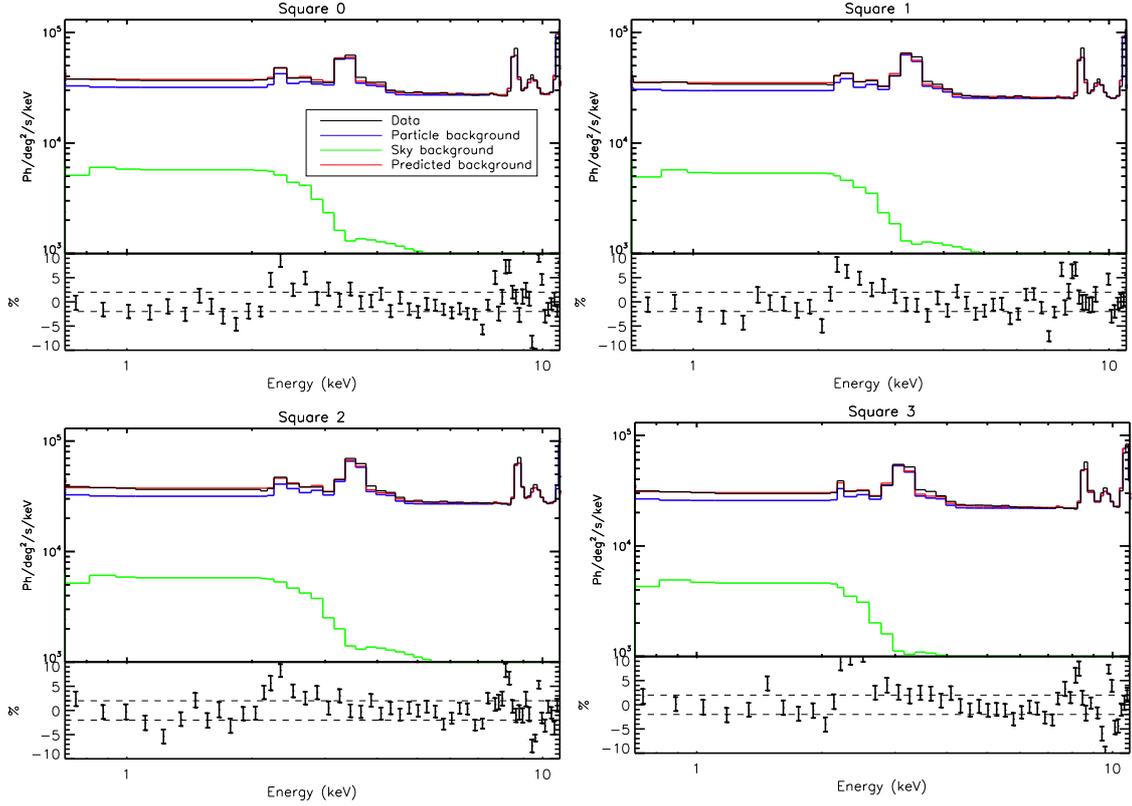}
  \end{center}
\caption{\textit{Spectra extracted from the four square regions shown in Fig. \ref{fig:mappa_cdfs}. 
Upper part of each panel: the black solid line represents data, the blue solid line the particle 
background, the green line the sky component, and the red line the total predicted background. 
Lower part of each panel: percentage variation of the data versus our background model defined as $100 \times (data/model-1)$. 
Dotted lines indicate the $\pm 2\%$ levels.}}
\label{fig:four_spectra}
\end{figure*}

To determine whether our model predicts the background spectrum inside the useful area of the FOV, we selected a square 
$\SI{12.5}{\arcmin} \times \SI{12.5}{\arcmin}$ 
area centred on the aim point and divided it into $2 \times 2$ square regions,  
shown with thick lines in Fig. \ref{fig:mappa_cdfs}.
The comparison of the spectrum extracted from these four regions and our background models is shown in Fig. \ref{fig:four_spectra}.
In all the four squares the predicted background is higher than the data in the $[1-2]\si{\kilo\electronvolt}$ band.
This systematic effect is always within the $2\%$ level, but inside this error there is hint of some sort 
of bias that causes the model to be higher than the data in this band. This might be partially related to a residual error in 
the effective area calibration and to a flattening of the power-law that is most likely related to the fact that we had better removed soft sources. 
We performed the fit of the gamma of the CXB power law on the spectrum extracted from the region defined by the combination of 
the four squares and the value decreased from $1.4$ to $1.2$. In this case, we obtain that the model 
predicts in the $[1-2]\si{\kilo\electronvolt}$ band a flux $\sim 0.4\%$ higher than the actual data.

\begin{figure}[!ht]
 \begin{center}
  \resizebox{\hsize}{!}{\includegraphics[bb=54 360 558 720]{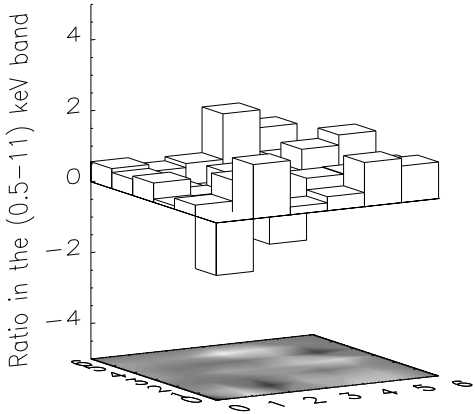}}
\end{center}
\caption{\textit{Percentage variation of the $[0.5-11] keV$ photon counts of the CDFS and the one predicted using the 
 proposed background model. Squares indicate the regions shown in Fig. \ref{fig:mappa_cdfs}.}}
\label{fig:ratio1}
\end{figure}

\begin{figure}[!ht]
 \begin{center}
  \resizebox{\hsize}{!}{\includegraphics[bb=54 360 558 720]{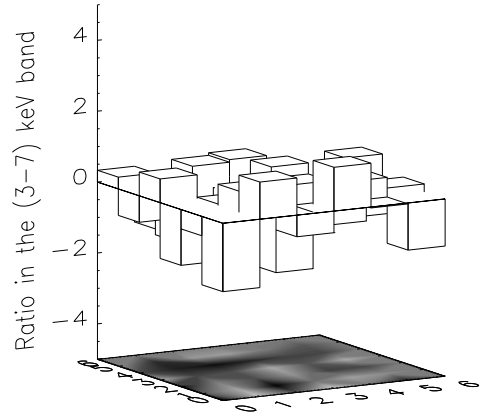}}
 \end{center}
\caption{\textit{Same as \ref{fig:ratio1}}, but in the $[3-7]keV$ energy band.}
\label{fig:ratio2}
\end{figure}

\begin{figure}[!ht]
 \begin{center}
  \resizebox{\hsize}{!}{\includegraphics[bb=54 360 558 720]{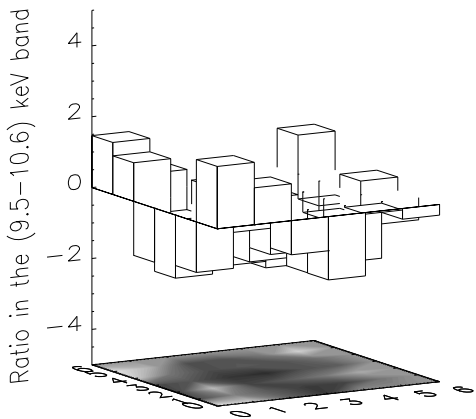}} 
 \end{center}
\caption{\textit{Same as \ref{fig:ratio1}, but in the $[9.5-10.6]keV$ energy band.}}
\label{fig:ratio3}
\end{figure}

To test the accuracy of our background model on smaller angular scales we further subdivided each previous square region into $3 \times 3$ square 
subregions. The new subregions are overlayed as thin squares in Fig. \ref{fig:mappa_cdfs}. Owing to the lower photon statistics available, for these 
subregions instead of the direct spectra comparison we computed the percentile variation of the detected and predicted photons in three 
different energy bands: $[0.5-11] \si{\kilo\electronvolt}$ (which may include 
some unresolved CXB), 
$[3-7] \si{\kilo\electronvolt}$ and $[9.5-10.6] \si{\kilo\electronvolt}$. In Figs. \ref{fig:ratio1} to \ref{fig:ratio3}, we show the results for the 
three bands. In the squares where the percentile variation has an absolute value higher than $2\%$ we show the percentile 
variation plus or minus, depending on whether the variation is higher than $2\%$ or lower than $-2\%$, respectively.

 These figures clearly show that our background model predicts the observed spectra with an accuracy better than $2\%$ in 
 any region in the continuum band. The accuracy of our model in 
the emission line regions is better than $5\%$ in any region. 
\section{Measuring CXB unresolved flux}\label{conclusioni}

The high accuracy of our particle background model easily allows us to measure the unresolved flux of CXB in the CDFS.
To do this, we selected two circular regions centred on (J$2000$) $RA=03:32:28.37$ $DEC=-27:48:10.1$ with a radius of $\SI{5}{\arcmin}$ and $\SI{8}{\arcmin}$. 
For the two circular regions we predict the total background, sky plus instrumental, by simply rescaling the 
values found from the fit described in Sec. \ref{risu_cdfs} for the area.  

In the flux estimation we excluded from our analysis the $[2-3]\si{\kilo\electronvolt}$ band and $[5.6-6.2]\si{\kilo\electronvolt}$ bands 
 to avoid contamination from emission lines.
We obtain a flux of $10.2^{+0.5}_{-0.4} 
\times 10^{-13}\si{\erg \per \square \centi\meter \deg}\si{\per \second}$ for the $[1-2]\si{\kilo\electronvolt}$
band and $(3.8 \pm 0.2) \times 10^{-12}\si{\erg \per \square \centi\meter \deg}\si{\per \second}$ for the $[2-8]\si{\kilo\electronvolt}$ band. Our 
results are perfectly consistent with those reported by \cite{mark_cxb}, but improve the accuracy by a factor larger than $3$.
\section{Summary and conclusions}\label{conclusion}

We studied the spatial and spectral variation of the ACIS-I Chandra particle background for the VF mode eight-year period D+E, starting in $2001$.

Using ACIS-stowed data for periods D+E, filtered using VF mode, we modelled the continuum as a sum of a power law plus an exponential,
the amplitude of
both following a gradient along the $y$ direction. Using the blank-sky observations available for the same period, we modelled 11 fluorescence lines. Six
of them are 
spatially variable and come in pairs of mother and daughter lines associated with the artefact of the CTI correction. 
In the future, this simple analytical model can be easily adjusted to match the upcoming background datasets.
We also showed that the best band to be used to normalize the particle background is $[9.5-10.6]\si{\kilo\electronvolt}$.
We demonstrated that our model is very stable over the detector, with an accuracy better than $2\%$ in the continuum and 
$5\%$ in the lines. Given these systematic errors and the statistics available, we were able to constrain the amplitude of the unresolved cosmic X-ray background from 
the CDFS with the unprecedented precision of $5\%$. 

One of the strengths of the Chandra observatory is its ability to resolve a large part of the CXB flux thanks to its unique angular resolution. 
Combining this 
ability with strong constraints on the systematic of its instrumental background should help us to best constrain the accuracy of imaging and 
spectral analyses
of faint extended sources. Among other aspects, the study of the outskirts of galaxy clusters, which is a well-known example for which
controlling the systematics of all background 
components is critical, will strongly benefit from our work (e.g. \citealt{cluster_out_1}, \citealt{cluster_out_2} or \citealt{cluster_out_3}).
\section*{Acknowledgments}
We thank Maxim Markevitch and the anonymous referee for their comments and suggestions, which significantly improved the quality of the paper.
\bibliographystyle{aa}
\bibliography{references_chandra_particle_AA_2014_23443}

\begin{table*}[!htbp]
\begin{center}
\caption{\textit{List of observation of the CDFS dataset}}
\label{tab:cdfs_data}
\resizebox{12.5cm}{!}{
\begin{tabular}{ccccc}
\hline 
\hline
 Obs ID   & R.A. &    Decl. & Exposure [$\si{\kilo\second}$] & Exposure after cleaning [$\si{\kilo\second}$] \\ 
\hline
 $8591$   & $03: 32: 28.20$ 	& $-27: 48: 36.00$ & $46.03$  & $45.32$    \\
 $8592$   & $03: 32: 28.20$ 	& $-27: 48: 36.00$ & $87.79$  & $86.64$    \\
 $8593$   & $03: 32: 28.20$ 	& $-27: 48: 36.00$ & $50.15$  & $47.63$    \\
 $8594$   & $03: 32: 28.20$ 	& $-27: 48: 36.00$ & $143.27$ & $140.63$   \\
 $8595$   & $03: 32: 28.20$ 	& $-27: 48: 36.00$ & $116.95$ & $114.90$   \\
 $8596$   & $03: 32: 28.20$ 	& $-27: 48: 36.00$ & $116.64$ & $ 113.58$  \\
 $8597$   & $03: 32: 28.20$ 	& $-27: 48: 36.00$ & $60.07$  & $59.02$    \\
 $9575$   & $03: 32: 28.20$ 	& $-27: 48: 36.00$ & $110.13$ & $106.8$    \\
 $9578$   & $03: 32: 28.20$ 	& $-27: 48: 36.00$ & $39.08$  & $38.01$    \\
 $9593$   & $03: 32: 28.20$ 	& $-27: 48: 36.00$ & $47.05$  & $43.23$    \\
 $9596$   & $03: 32: 28.20$ 	& $-27: 48: 36.00$ & $113.37$ & $113.79$   \\
 $9718$   & $03: 32: 28.20$ 	& $-27: 48: 36.00$ & $50.03$  & $49.35$    \\
 $12043$  & $03: 32: 28.80$ 	& $-27: 48: 23.00$ & $131.29$ & $128.55$   \\
 $12044$  & $03: 32: 28.80$ 	& $-27: 48: 23.00$ & $100.84$ & $99.52$    \\
 $12055$  & $03: 32: 28.80$ 	& $-27: 48: 23.00$ & $81.75$  & $79.91$    \\
 $12123$  & $03: 32: 28.80$ 	& $-27: 48: 23.00$ & $25.12$  & $24.50$    \\
 $12128$  & $03: 32: 28.80$ 	& $-27: 48: 23.00$ & $23.1$   & $22.80$    \\
 $12213$  & $03: 32: 28.80$ 	& $-27: 48: 23.00$ & $62.1$   & $61.03$    \\
 $12045$  & $03: 32: 28.80$ 	& $-27: 48: 23.00$ & $101.04$ & $99.46$    \\
 $12046$  & $03: 32: 28.80$ 	& $-27: 48: 23.00$ & $79.06$  & $77.51$    \\
 $12047$  & $03: 32: 28.80$ 	& $-27: 48: 23.00$ & $10.28$  & $10.14$    \\
 $12048$  & $03: 32: 28.80$ 	& $-27: 48: 23.00$ & $139.93$ & $137.85$   \\
 $12129$  & $03: 32: 28.80$ 	& $-27: 48: 23.00$ & $78.16$  & $76.63$    \\
 $12135$  & $03: 32: 28.80$ 	& $-27: 48: 23.00$ & $63.36$  & $62.27$    \\
 $12137$  & $03: 32: 28.80$ 	& $-27: 48: 23.00$ & $94.01$  & $91.97$    \\ 
 $12138$  & $03: 32: 28.80$ 	& $-27: 48: 23.00$ & $39.04$  & $38.53$    \\
 $12049$  & $03: 32: 28.80$ 	& $-27: 48: 23.00$ & $88.09$  & $86.43$    \\
 $12050$  & $03: 32: 28.80$ 	& $-27: 48: 23.00$ & $30.05$  & $29.65$    \\
 $12051$  & $03: 32: 28.80$ 	& $-27: 48: 23.00$ & $58.04$  & $56.45$    \\
 $12218$  & $03: 32: 28.80$ 	& $-27: 48: 23.00$ & $89.14$  & $87.46$    \\
 $12219$  & $03: 32: 28.80$ 	& $-27: 48: 23.00$ & $34.11$  & $33.66$    \\
 $12220$  & $03: 32: 28.80$ 	& $-27: 48: 23.00$ & $48.77$  & $47.86$    \\
 $12222$  & $03: 32: 28.80$ 	& $-27: 48: 23.00$ & $31.05$  & $30.64$    \\
 $12223$  & $03: 32: 28.80$ 	& $-27: 48: 23.00$ & $102.04$ & $99.68$    \\
 $12052$  & $03: 32: 28.80$ 	& $-27: 48: 23.00$ & $111.88$ & $110.41$   \\
 $12053$  & $03: 32: 28.80$ 	& $-27: 48: 23.00$ & $69.01$  & $67.85$    \\
 $12054$  & $03: 32: 28.80$ 	& $-27: 48: 23.00$ & $61.81$  & $60.75$    \\
 $12227$  & $03: 32: 28.80$ 	& $-27: 48: 23.00$ & $55.04$  & $54.05$    \\
 $12230$  & $03: 32: 28.80$ 	& $-27: 48: 23.00$ & $34.25$  & $33.55$    \\
 $12231$  & $03: 32: 28.80$ 	& $-27: 48: 23.00$ & $25.05$  & $24.72$    \\
 $12232$  & $03: 32: 28.80$ 	& $-27: 48: 23.00$ & $32.89$  & $32.89$    \\
 $12233$  & $03: 32: 28.80$ 	& $-27: 48: 23.00$ & $35.57$  & $35.32$    \\
 $12234$  & $03: 32: 28.80$ 	& $-27: 48: 23.00$ & $49.15$  & $49.15$    \\
          &                    &                &          &            \\  
Total Exp &                    &                &          &  $3024.36$ \\ 
\hline
\end{tabular}
}
\end{center}
\end{table*}
\clearpage

\clearpage
\begin{table*}[!ht]
 \caption{Best-fit values for the varying lines of the three MDSB of the CCD $0$ (equation \ref{eq:mother_and_daughter}).}
 \label{tab:ccd0_par}
  \begin{center}
   \begin{tabular}{cccc}
 \hline
 \hline
Parameter                               &  $\gamma^{0}$              & $\gamma^{1}$             & $\gamma^{2}$ \\
\hline
$B_{1} (\si{\kilo\electronvolt})$       &$      2.169$               &$0$                       &$0$\\
$Q_{1}   (\si{\kilo\electronvolt})$     &$    5.295 \times 10^{-2} $ &$0$                       & $0$       \\
$\Phi_{1}$                              & $    2.644 \times 10^{-2}$ &$ -8.938 \times 10^{-7}$  &$  3.988 \times 10^{-9}$\\
 $\delta_{1} (\si{\kilo\electronvolt})$ &$    8.477 \times 10^{-2}$  &$  3.954 \times 10^{-4}$  &$0$\\
$\nu{1} (\si{\kilo\electronvolt})$      &$    4.974 \times 10^{-2}$  &$ -2.263 \times 10^{-5}$  &$  5.469 \times 10^{-8}$\\
$\phi{1}$                               &$   1.451 \times 10^{-3}$   &$  7.868 \times 10^{-6}$  &$0$\\
\hline
$B_{2} (\si{\kilo\electronvolt})$       &$      7.498$               &$ 0$  &$0$\\
$Q_{2}   (\si{\kilo\electronvolt})$     &$    4.063 \times 10^{-2}$  &$0$                       & $0$       \\
$\Phi_{2}$                              &$    2.928 \times 10^{-2} $ &$  5.714 \times 10^{-5}$ &$ -5.318 \times 10^{-8}$\\
 $\delta_{2} (\si{\kilo\electronvolt})$ &$     1.055 \times 10^{-1}$ &$  7.247 \times 10^{-4}$ &$0$\\
$\nu{2} (\si{\kilo\electronvolt})$      &$    8.973 \times 10^{-2}$  &$ -3.426 \times 10^{-5}$&$  1.029 \times 10^{-7}$\\
$\phi{2}$                               &$  -6.053 \times 10^{-3}$   &$  1.653 \times 10^{5}$  &$0$\\
\hline
$B_{3} (\si{\kilo\electronvolt})$       &$      9.727$               &$0$&$0$\\
$Q_{3}   (\si{\kilo\electronvolt})$     &$    4.567 \times 10^{-2}$  &$0$                     & $0$       \\
$\Phi_{3}$                              &$     1.085 \times 10^{-1}$ &$  1.922 \times 10^{-5}$&$ -2.491 \times 10^{-8}$\\
 $\delta_{3} (\si{\kilo\electronvolt})$ &$     1.593 \times 10^{-1}$ &$  8.308 \times 10^{-4}$&$0$\\
$\nu{3} (\si{\kilo\electronvolt})$      &$     1.097 \times 10^{-1} $&$ -7.152 \times 10^{-5}$&$  1.493 \times 10^{-7}$\\
$\phi{3}$                               &$  -3.319 \times 10^{-3}$   &$  3.574 \times 10^{-5}$&$0$\\
   \end{tabular}
  \end{center}
 \end{table*}
\begin{table*}[!ht]
 \caption{Same as Table \ref{tab:ccd0_par}, but for CCD $1$.}
 \label{tab:ccd1_par}
  \begin{center}
   \begin{tabular}{cccc}
 \hline
 \hline
Parameter                               &  $\gamma^{0}$              &$\gamma^{1}$             & $\gamma^{2}$ \\
\hline
$B_{1} (\si{\kilo\electronvolt})$       &$      2.166$               &$0$ &$0$\\
$Q_{1}   (\si{\kilo\electronvolt})$     &$    5.295\times 10^{-2}$   &$0$                      & $0$       \\
$\Phi_{1}$                              &$    2.898\times 10^{-2}$   &$ -6.182 \times 10^{-6}$ &$  7.427 10^{-9}$\\
 $\delta_{1} (\si{\kilo\electronvolt})$ &$    9.332\times 10^{-2}$   &$  3.925\times 10^{-4}$  &$0$\\
$\nu{1} (\si{\kilo\electronvolt})$      &$    5.184\times 10^{-2}$   &$ -4.094 \times 10^{-5}$ &$  7.352 \times 10^{-8}$\\
$\phi{1}$                               &$   -1.728\times 10^{-3}$   &$  1.222 \times 10^{-5}$ &$0$\\
\hline
$B_{2} (\si{\kilo\electronvolt})$       &$      7.481$               &$ 0$ &$0$\\
$Q_{2}   (\si{\kilo\electronvolt})$     &$    4.633\times 10^{-2}$   &$0$                      & $0$       \\
$\Phi_{2}$                              &$    3.363\times 10^{-2}$   &$  3.113 \times 10^{-5}$ &$ -2.631e \times 10^{-8}$\\
 $\delta_{2} (\si{\kilo\electronvolt})$ &$     1.289\times 10^{-1}$  &$  7.279 \times 10^{-4}$ &$0$\\
$\nu{2} (\si{\kilo\electronvolt})$      &$    9.582\times 10^{-2}$   &$ -5.605 \times 10^{-5}$ &$  1.184 \times 10^{-7}$\\
$\phi{2}$                               &$  -6.128\times 10^{-3}$    &$  2.034 \times 10^{-5}$ &$0$\\
\hline
$B_{3} (\si{\kilo\electronvolt})$       &$      9.712$               &$ 0$ &$0$\\
$Q_{3}   (\si{\kilo\electronvolt})$     &$    4.567\times 10^{-2}$   &$0$                      & $0$       \\
$\Phi_{3}$                              &$     1.140\times 10^{-1}$  &$  8.515 \times 10^{-6}$ &$ -1.948 \times 10^{-8}$\\
 $\delta_{3} (\si{\kilo\electronvolt})$ &$     1.392\times 10^{-1}$  &$  8.199\times 10^{-4}$  &$0$\\
$\nu{3} (\si{\kilo\electronvolt})$      &$     1.158\times 10^{-1}$  &$ -8.769 \times 10^{-5}$ &$  1.585 \times 10^{-7}$\\
$\phi{3}$                               &$   1.049\times 10^{-3}$    &$  3.003 \times 10^{-5}$ &$0$\\
   \end{tabular}
  \end{center}
 \end{table*}
\clearpage

\begin{table*}[!ht]
 \caption{Same as Table \ref{tab:ccd0_par}, but for CCD $2$.}
 \label{tab:ccd2_par}
  \begin{center}
   \begin{tabular}{cccc}
 \hline
 \hline 
Parameter                                &  $\gamma^{0}$              & $\gamma^{1}$           & $\gamma^{2}$ \\
\hline
$B_{1} (\si{\kilo\electronvolt})$        &$      2.157$               &$ 0$                    &$0$\\
$Q_{1}   (\si{\kilo\electronvolt})$      &$    5.295\times 10^{-2}$   &$0$                     & $0$       \\
$\Phi_{1}$                               &$    2.570\times 10^{-2}$   &$ -5.391 \times 10^{-7}$&$  2.627 \times 10^{-9}$\\
 $\delta_{1} (\si{\kilo\electronvolt})$  &$    5.459\times 10^{-2}$   &$  5.039 \times 10^{-4}$&$0$\\
$\nu{1} (\si{\kilo\electronvolt})$       &$    6.118\times 10^{-2}$   &$ -3.738 \times 10^{-5}$&$  7.397 \times 10^{-8}$\\
$\phi{1}$                                &$ -1.832\times 10^{-4}$     &$  1.130 \times 10^{-5}$&$0$\\
\hline
$B_{2} (\si{\kilo\electronvolt})$        &$      7.504$               &$ 0$                    & $0$\\
$Q_{2}   (\si{\kilo\electronvolt})$      &$    4.633\times 10^{-2}$   &$0$                     & $0$       \\
$\Phi_{2}$                               &$    3.344\times 10^{-2}$   &$  5.135 \times 10^{-5}$&$ -4.418 \times 10^{-8}$\\
 $\delta_{2} (\si{\kilo\electronvolt})$  &$    3.348\times 10^{-2}$   &$  9.058 \times 10^{-4}$&$0$\\
$\nu{2} (\si{\kilo\electronvolt})$       &$     1.097\times 10^{-1}$  &$ -3.611 \times 10^{-5}$&$  1.103 \times 10^{-7}$\\
$\phi{2}$                                &$  -1.168\times 10^{-3}$    &$  1.102 \times 10^{-5}$&$0$\\
\hline
$B_{3} (\si{\kilo\electronvolt})$    &$      9.746$                   &$ 0$&$0$\\
$Q_{3}   (\si{\kilo\electronvolt})$      &$    4.567\times 10^{-2}$   &$0$                     & $0$       \\
$\Phi_{3}$                               &$     1.129\times 10^{-1}$  &$ -1.575 \times 10^{-5}$&$  3.478 \times 10^{-9}$\\
 $\delta_{3} (\si{\kilo\electronvolt})$  &$    9.727\times 10^{-2}$   &$  1.039 \times 10^{-3}$&$0$\\
$\nu{3} (\si{\kilo\electronvolt})$       &$     1.329\times 10^{-1}$  &$ -8.840 \times 10^{-5}$&$  1.795 \times 10^{-7}$\\
$\phi{3}$                                &$   2.727\times 10^{-3}$    &$  3.251 \times 10^{-5}$&$0$\\

   \end{tabular}
  \end{center}
 \end{table*}
\begin{table*}[!ht]
 \caption{Same as Table \ref{tab:ccd0_par}, but for CCD $3$.}
 \label{tab:ccd3_par}
  \begin{center}
   \begin{tabular}{cccc}
 \hline
 \hline
Parameter                               &  $\gamma^{0}$                & $\gamma^{1}$           & $\gamma^{2}$ \\
\hline
$B_{1} (\si{\kilo\electronvolt})$       &$      2.164$                 &$0$ &$0$\\
$Q_{1}   (\si{\kilo\electronvolt})$     &$    5.295 \times 10^{-2}$    &$0$                     & $0$       \\
$\Phi_{1}$                              &$    2.512 \times 10^{-2}$    &$  3.335 \times 10^{-6}$&$  1.143 \times 10^{-9}$\\
 $\delta_{1} (\si{\kilo\electronvolt})$ &$    9.199 \times 10^{-2}$    &$  4.939 \times 10^{-4}$&$0$\\
$\nu{1} (\si{\kilo\electronvolt})$      &$    5.860 \times 10^{-2}$    &$ -2.661 \times 10^{-5}$&$  6.403 \times 10^{-8}$\\
$\phi{1}$                               &$  -1.974 \times 10^{-3}$     &$  1.548 \times 10^{-5}$&$0$\\
\hline
$B_{2} (\si{\kilo\electronvolt})$       &$      7.482$                 &$0$                     &$0$\\
$Q_{2}   (\si{\kilo\electronvolt})$     &$    4.633 \times 10^{-2}$    &$0$                     & $0$       \\
$\Phi_{2}$                              &$    3.843 \times 10^{-2}$    &$  1.610 \times 10^{-5}$&$ -1.047 \times 10^{-8}$\\
 $\delta_{2} (\si{\kilo\electronvolt})$ &$    1.500\times 10^{-1}$     &$  8.922 \times 10^{-4}$&$0$\\
$\nu{2} (\si{\kilo\electronvolt})$      &$    1.121\times 10^{-1}$     &$ -4.785 \times 10^{-5}$&$  1.173 \times 10^{-7}$\\
$\phi{2}$                               &$  -1.692 \times 10^{-3}$     &$  1.526 \times 10^{-5}$&$0$\\
\hline
$B_{3} (\si{\kilo\electronvolt})$       &$      9.718$                 &$ 0$&$0$\\
$Q_{3}   (\si{\kilo\electronvolt})$     &$    4.567 \times 10^{-2}$    &$0$                     & $0$       \\
$\Phi_{3}$                              &$     1.138 \times 10^{-1}$   &$ -9.232 \times 10^{-6}$&$ -5.671 \times 10^{-9}$\\
 $\delta_{3} (\si{\kilo\electronvolt})$ &$     1.252 \times 10^{-1}$   &$  1.071 \times 10^{-3}$&$0$\\
$\nu{3} (\si{\kilo\electronvolt})$      &$     1.432 \times 10^{-1}$   &$ -1.311 \times 10^{-4}$&$  2.177 \times 10^{-7}$\\
$\phi{3}$                               &$  -1.808 \times 10^{-3}$     &$  3.773 \times 10^{-5}$&$0$\\

   \end{tabular}
  \end{center}
 \end{table*}

\end{document}